\documentclass[a4paper]{article}
\usepackage{amssymb}
\usepackage{graphicx}
\usepackage{amssymb}
\usepackage{amsmath}
\usepackage{xcolor}
\usepackage[cm]{fullpage}
\usepackage{upgreek}
\usepackage{float}
\usepackage{siunitx}
\usepackage[labelfont=bf]{caption}
\usepackage{tocstyle}
\usetocstyle{KOMAlike}

\newcommand{\figureref}[1]{Figure~#1}

\begin{document}

\title{\textbf{Quantum simulation of a Fermi-Hubbard model using a semiconductor quantum dot array}}

\author{T. Hensgens$^{1}$, T. Fujita$^{1}$, L. Janssen$^{1}$, Xiao Li$^{2}$, C.~J. Van Diepen$^{3}$, C. Reichl$^{4}$\\W. Wegscheider$^{4}$, S. Das Sarma$^{2}$ \& L.~M.~K. Vandersypen$^{1}$\footnote{email: l.m.k.vandersypen@tudelft.nl}}
\maketitle

\begin{enumerate}
\centering
 \item QuTech and Kavli Institute of Nanoscience, TU Delft, 2600 GA Delft, The Netherlands
 \item Condensed Matter Theory Center and Joint Quantum Institute, University of Maryland, College Park, Maryland 20742, USA
 \item QuTech and Netherlands Organization for Applied Scientific Research (TNO), 2600 AD Delft, The Netherlands
 \item Solid State Physics Laboratory, ETH Z\"{u}rich, 8093 Z\"{u}rich, Switzerland
\end{enumerate}

\section*{\centering Abstract} 
Interacting fermions on a lattice can develop strong quantum correlations, which lie at the heart of the classical intractability of many exotic phases of matter \cite{Imada1998,Lee2006,Balents2010,Anderson2013}.
Seminal efforts are underway in the control of artificial quantum systems, that can be made to emulate the underlying Fermi-Hubbard models \cite{Joerdens2008,Cirac2012,Tanese2013,Parsons2016,Boll2016,Cheuk2016,Mazurenko2016}.
Electrostatically confined conduction band electrons define interacting quantum coherent spin and charge degrees of freedom that allow all-electrical pure-state initialisation and readily adhere to an engineerable Fermi-Hubbard Hamiltonian \cite{Manousakis2002,Gaudreau2006,Byrnes2008,Yang2011,Barthelemy2013,Loss1998a,Petta2005,Koppens2006,Petersson2010,Baart2015,Martins2016,Reed2016}. Until now, however, the substantial electrostatic disorder inherent to solid state has made attempts at emulating Fermi-Hubbard physics on solid-state platforms few and far between \cite{Singha2011,Salfi2016}.
Here, we show that for gate-defined quantum dots, this disorder can be suppressed in a controlled manner.
Novel insights and a newly developed semi-automated and scalable toolbox allow us to homogeneously and independently dial in the electron filling and nearest-neighbour tunnel coupling. Bringing these ideas and tools to fruition, we realize the first detailed characterization of the collective Coulomb blockade transition \cite{Stafford1994}, which is the finite-size analogue of the interaction-driven Mott metal-to-insulator transition \cite{Imada1998}.
As automation and device fabrication of semiconductor quantum dots continue to improve, the ideas presented here show how quantum dots can be used to investigate the physics of ever more complex many-body states.

\section*{\centering Introduction}
The Fermi-Hubbard model, which describes interacting electrons on a lattice of tunnel-coupled sites, is central to many long-standing problems in low-dimensional condensed-matter physics, in topics ranging from high-$T_c$ superconductivity to electronic spin liquids \cite{Imada1998,Lee2006,Balents2010,Anderson2013}. Although the potential of such correlated-electron phases for realizing novel electronic and magnetic properties has prompted quantum simulation efforts across multiple platforms \cite{Joerdens2008,Cirac2012,Tanese2013,Parsons2016,Boll2016,Cheuk2016,Singha2011,Salfi2016}, experimental correlations are often limited in span and strength due to the residual entropy of the initialized system \cite{Parsons2016,Boll2016,Cheuk2016}. Furthermore, scaling to similarly homogeneous but larger system sizes is not always straightforward \cite{Joerdens2008,Tanese2013,Parsons2016,Boll2016,Cheuk2016,Mazurenko2016,Salfi2016}.
Semiconductor quantum dots form a scalable platform that is naturally described by a Fermi-Hubbard model in the low-temperature, strong-interaction regime, when cooled down to dilution temperatures \cite{Manousakis2002,Gaudreau2006,Yang2011,Byrnes2008,Barthelemy2013}. As such, pure state initialization of highly-entangled states is possible even without the use of adiabatic initialization schemes \cite{Farooq2015}. Coherent evolution of excitations in charge and spin can span many sites, as, contrary to what might be expected, dissipation and decoherence rates induced by electromagnetic noise can be made $>20$ times smaller than the relevant coupling energies \cite{Petersson2010,Martins2016,Reed2016}. Furthermore, local control and read-out of both charge and spin degrees of freedom have become matured areas of research, given the large ongoing effort of using quantum dots as a platform for quantum information processing \cite{Loss1998a,Petta2005,Koppens2006,Petersson2010,Baart2015,Martins2016,Reed2016}.
In particular, excellent control of small on-site energy differences \cite{Petersson2010,Stehlik2015} or tunnel couplings \cite{Martins2016,Reed2016} has been shown at specific values of electron filling and tuning. Quantum simulation experiments can leverage many of these developments, trading off some of the experimental difficulties involved in full coherent control for ease of scaling.
Until now, however, calibration routines for quantum dots have been quite inefficient and limited in scope. As such, the effective control of larger parameter spaces as well as the calibration of larger samples seem like insurmountable obstacles.
What has been lacking, thus, is an efficient and scalable control paradigm for Hamiltonian engineering that extends to the collective Fermi-Hubbard parameter regimes well beyond those required for qubit operation \cite{Kouwenhoven1990,Livermore1996,Lee2000,Wei2016} and described merely by (local) spin excitations.\\

\noindent In this Letter we demonstrate the simulation of Fermi-Hubbard physics using semiconductor quantum dots. We describe an experimental toolbox, validated by direct numerical simulations, that allows for the independent tuning of filling and tunnel coupling as well as the measurement of all interaction energies, and employ it to map out the accessible parameter space of a triple quantum dot device with unprecendented detail and precision. As the tunnel couplings are homogeneously increased, we witness the delocalization transition between isolated Coulomb blockade and collective Coulomb blockade, a finite-size analogue of the interaction-driven Mott transition.\\

\section*{\centering Methods and Results}

\noindent The one-dimensional quantum dot array is electrostatically defined using voltages applied to gate electrodes fabricated on the surface of a GaAs/AlGaAs heterostructure (\figureref{1}), that selectively deplete regions of the 85-nm-deep two-dimensional electron gas (2DEG) underneath. The outermost dots can be (un)loaded from Fermi reservoirs on the sides, which have an effective electron temperature of 72.2(5) mK (6.26(5) $\upmu$eV). The three gates at the top are used to define a sensing-dot channel, the conductance of which is sensitive to changes in the charge state of the array and is directly read out using radio-frequency reflectometry.\\

\begin{figure*}[h!]
	\centering
 	\includegraphics{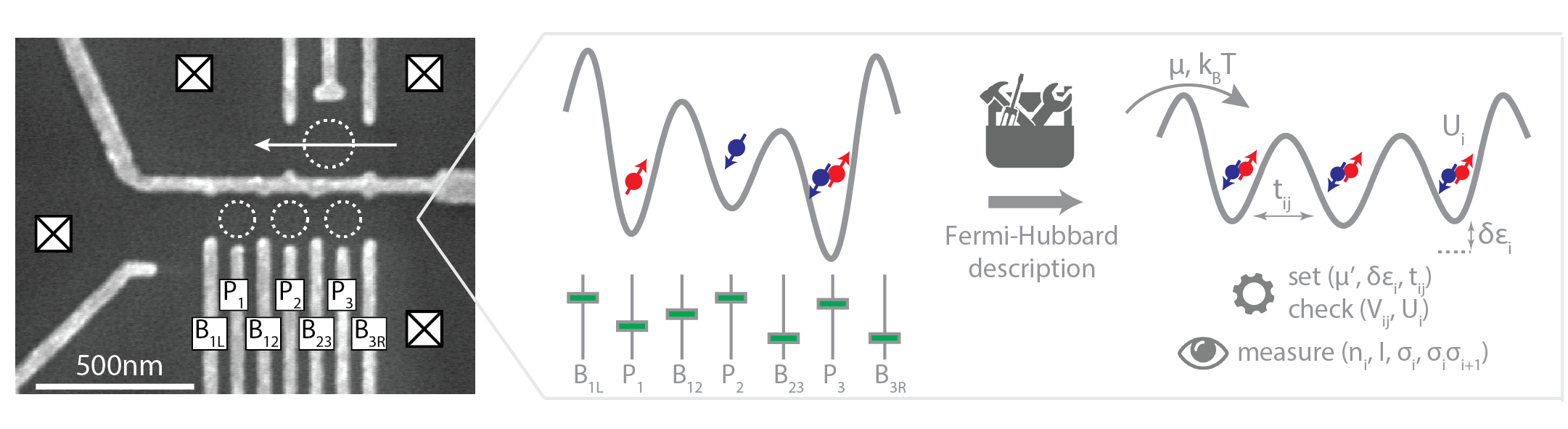}
 	\caption{\textbf{Gate-defined quantum dot array as a platform for quantum simulations of the Fermi-Hubbard model.}
Electron micrograph of a sample nominally identical to the one used for the measurements. The bottom three circles indicate the triple dot array, whose Hamiltonian parameters derive from the local potential landscape controlled by the seven bottom-most gates ($B_{1L}$ to $B_{3R}$). The top circle and arrow indicate the sensing dot channel, the radio-frequency reflectance of which is monitored to enable real-time charge sensing.  Crossed squares indicate distinct Fermi reservoirs that are contacted using ohmic contacts. We describe a toolbox that allows for the control of the quantum dot array at the level of the microscopic Fermi-Hubbard model. In particular, it allows for the independent calibration of \{$\mu',\delta\epsilon_i,t_{ij}$\} and the measurement of the Coulomb interaction terms \{$V_{ij},U_i$\}. Measurable observables for quantum dots include both local charge occupation and global charge transport as well as local spin degrees of freedom and nearest-neighbour singlet-triplet spin correlations (through spin-to-charge conversion protocols \cite{Petta2005,Baart2015}). }
 	\label{fig:virtgates}
 \end{figure*} 

\noindent The control of Fermi-Hubbard model parameters is achieved by modulation of the potential landscape in the 2DEG using the seven bottom-most gate electrodes (\figureref{1}). These gates come in two flavours. Plunger gates $P_i$ are designed to tune the single-particle energy offsets $\epsilon_i$ of individual dots $i$, allowing us to set an overall chemical potential $\mu'=\langle \epsilon_i \rangle$ and add site-specific detuning terms $\delta\epsilon_i$ (see Supplementary Section I). Barrier gates $B_{ij}$ allow for the modulation of tunnel couplings $t_{ij}$ between the $i$th and $j$th dot or $\Gamma_{ij}$ between the $i$th dot and $j$th Fermi reservoir, respectively. The interaction energies are determined by the potential landscape realized to achieve this set \{$\mu',\delta\epsilon_i$,$t_{ij}$,$\Gamma_{ij}$\}, and comprise of the on-site Coulomb interaction terms $U_i$ and inter-site Coulomb interaction terms $V_{ij}$. With each dot filled to an even number of electrons, we can describe the addition of the next two electrons per dot within an effective single-band extended Hubbard picture \cite{Wang2011}, using site-and-spin-specific electronic creation and annihilation operators $c^{\dag}_{i\sigma}$ and $c_{i\sigma}$ and dot occupations $n_{i}=\sum_{\sigma}{c^{\dag}_{i\sigma}c_{i\sigma}}$:\\

\begin{equation}
H = - \sum_i{\epsilon_i n_i} - \sum_{<i,j>,\sigma}{t_{ij}(c^{\dag}_{i\sigma}c_{j\sigma}+\mathrm{h.c.})} + \sum_{i}{\frac{U_i}{2}n_i(n_i-1)} + \sum_{i,j}{V_{ij}n_in_j}.
\label{eq:Hubbarmodel}
\end{equation}\\

\begin{figure*}[h!]
	\centering
 	\includegraphics{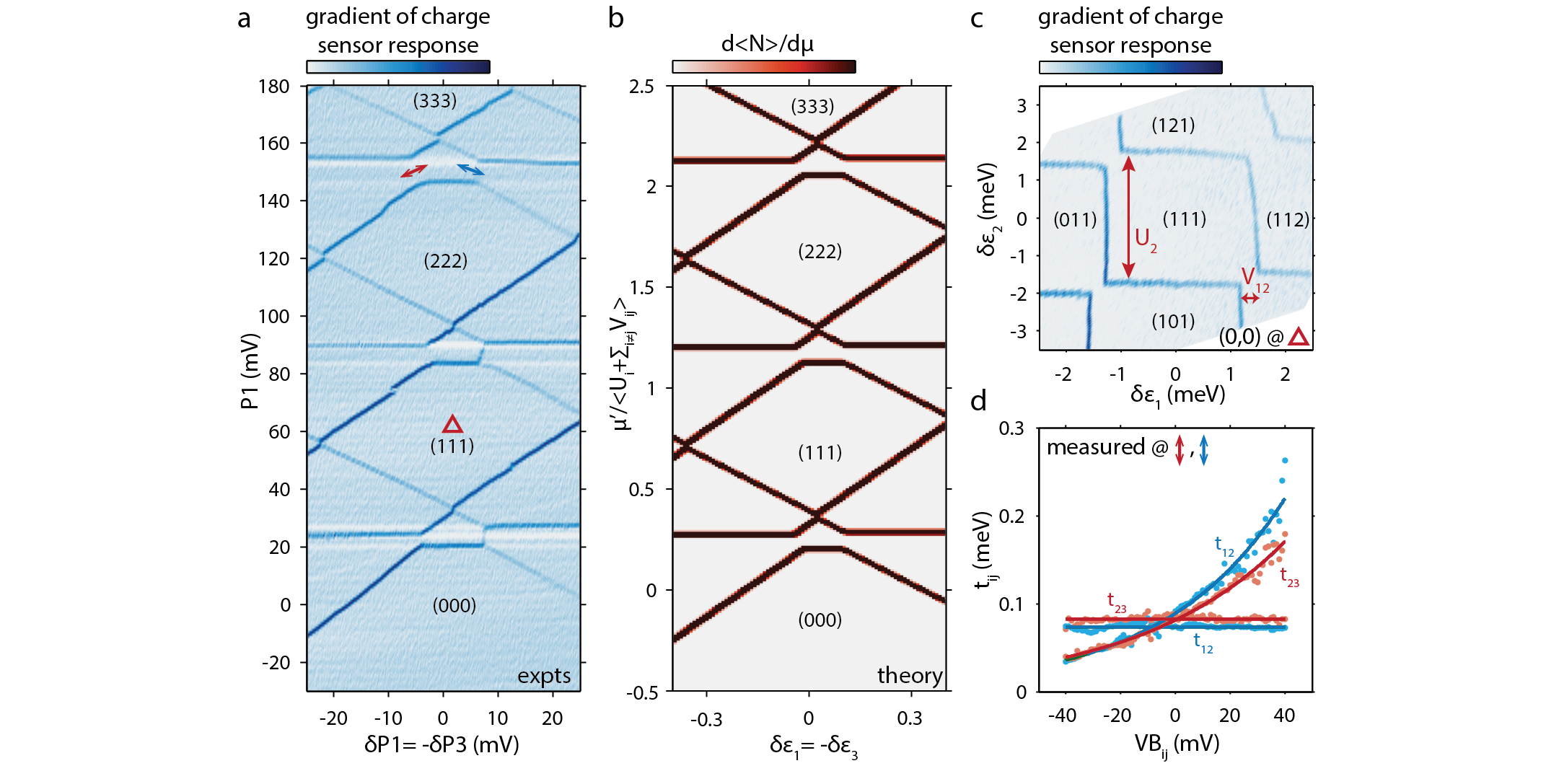}
 	\caption{\textbf{Hamiltonian engineering using a scalable toolbox of local control and measurements.} \textbf{a} Charge stability diagram showing uniform filling of the array of up to three electrons per dot in the vertical direction, using a combination of all seven gates (only $P_1$ values are shown) that equally sweeps the local fillings  $n_i$ while keeping the tunnel couplings between dots and to the reservoirs nominally identical. The dark lines arise from steps in the charge detector conductance, indicating a transition in the number of electrons on one of the dots. \textbf{b} Theoretical charge stability spectrum of a triple-quantum-dot system in the classical limit ($t=0$) exchanging particles with a reservoir at $U/k_BT=300$, analogous to the measurement in \textbf{a}. \textbf{c} As we focus on relevant sections of the charge-stability diagram of the array, we calibrate all relative cross-capacitances of the seven-gate, three dot-system, allowing for the deterministic setting of changes in $\epsilon_i$ and subsequent measurement of on-site and inter-site Coulomb couplings. Furthermore, it allows for the definition of virtual gates that influence the tunnel couplings without changing the $\epsilon_i$, fixing the location in the charge stability diagram and making automated repeated measurements possible. \textbf{d} Measurements of both tunnel couplings as a function of two linear combinations of gate voltages that keep either $t_{12}$ or $t_{23}$ (the full line denotes the average value) as well as the three on-site energies $\epsilon_i$ constant whilst increasing $t_{23}$ or $t_{12}$ (an exponential fit to $\alpha \exp (VB_{ij} / \beta)$ is shown), respectively. Individual tunnel coupling measurements are taken at a rate of roughly 1 Hz. Text in brackets denote the dominant charge states in the many-body eigenstate.}
 	\label{fig:virtgates}
 \end{figure*}

\noindent In practice, both $P_i$ and $B_{ij}$ gates exhibit cross-talk to all the $\epsilon_i$ and $t_{ij}$ (with smaller effects on $U_i$ and $V_{ij}$), and in addition must compensate for initial disorder. Setting Hamiltonian parameters experimentally therefore requires carefully chosen linear combinations of gate voltages. This idea is employed regularly in spin qubit experiments in order to change the on-site energies $\epsilon_i$ deterministically over small ranges \cite{Oosterkamp1998}, but here we go further in important ways. Our experimental toolbox uses linear combinations of gate voltage changes \{$P_i,B_{ij}$\} for the independent control of the Fermi-Hubbard parameters \{$\mu',\delta\epsilon_i$,$t_{ij}$\} to within several $k_BT$ and over a wide range of fillings and tunnel couplings. \figureref{2a-b} shows the filling of the array with up to nine electrons, three electrons per dot, while keeping the inter-dot tunnelling terms small ($t_{ij}<V_{ij}<U_i$) and the tunnel couplings to the reservoirs roughly constant (see Supplementary Section I). The three different slopes of the charge transition lines correspond to filling of each of the three dots. \figureref{2c} shows the addition of site-specific detuning, and \figureref{2d} the independent control of both tunnel couplings. To achieve this level of control required several new insights. First, we measure the cross-capacitance terms between the three dots and seven gates repeatedly, allowing for the direct definition of virtual $\delta\epsilon_i$ gates that are accurate over a range of several meV (see Supplementary section II). This in turn allows us to compensate for the effect that barrier gates have on dot detunings, reducing the problem of finding orthogonal virtual $t_{ij}$ gates to a simple cross-talk measurement. For the vertical $\mu'$ axis shown in \figureref{2a}, we first note that homogeneous filling requires non-homogeneous changes in the $\epsilon_i$, as the dots have to each overcome a different sum of local interaction energies $U_i+\sum_{i \neq j} V_{ij}$. In addition, we must add the compensated barrier gates described above to counter both the effect that changing plunger gate voltages and the higher wave function overlap of higher electron fillings have on the tunnel couplings.\\

\noindent Having filled the array with a given number of electrons, we can quantitatively characterize the various parameters in the Fermi-Hubbard model directly from relevant feature sizes in the charge stability diagram as we detune away from uniform filling.
The spacing between charge addition lines of half-filled dot levels yields the on-site Coulomb interaction term $U_i$, whereas the displacement of single charge addition lines upon filling another dot yields their inter-site Coulomb coupling $V_{ij}$ (see \figureref{2c} and Supplementary Section III for details). 
Finally, from the width of a polarization line, where an electron moves between adjacent sites, we can extract the tunnel coupling $t_{ij}$. We implement an iterative tuning process that leverages both knowledge of the cross-capacitances as well as the automation of the tunnel coupling measurement technique (see arrows in \figureref{2a}, \figureref{2d} and Supplementary Section IV for details).\\

\noindent We showcase the potential of well-controlled quantum dot arrays to emulate Fermi-Hubbard physics by employing this newly developed toolbox for the realization of collective Coulomb blockade (CCB) physics, validating the results through direct numerical Fermi-Hubbard model calculations.
Coulomb blockade (CB) is a purely classical effect that arises from the finite charging energies of each individual quantum dot, where the charge excitations at half filling are gapped out, analogous to the Mott gap (\figureref{3a}). When quantum tunneling effects between sites are turned on, however, a much richer phase diagram appears. The CB of individual dots is destroyed as the degeneracy of the peaks in the equilibrium charge addition spectrum (see Supplementary Section V) is lifted and broadened into minibands, giving way to collective Coulomb blockade \cite{Stafford1994} (\figureref{3a}). As tunnel couplings continue to increase relative to local charging energies this gap will vanish in the thermodynamic limit, giving rise to a metallic state. The CCB physics is best described by the equilibrium electron addition spectrum as a function of filling and tunnel coupling, the two main experimental control parameters of our quantum dot array.\\

\begin{figure*}[h!]
	\centering
 	\includegraphics{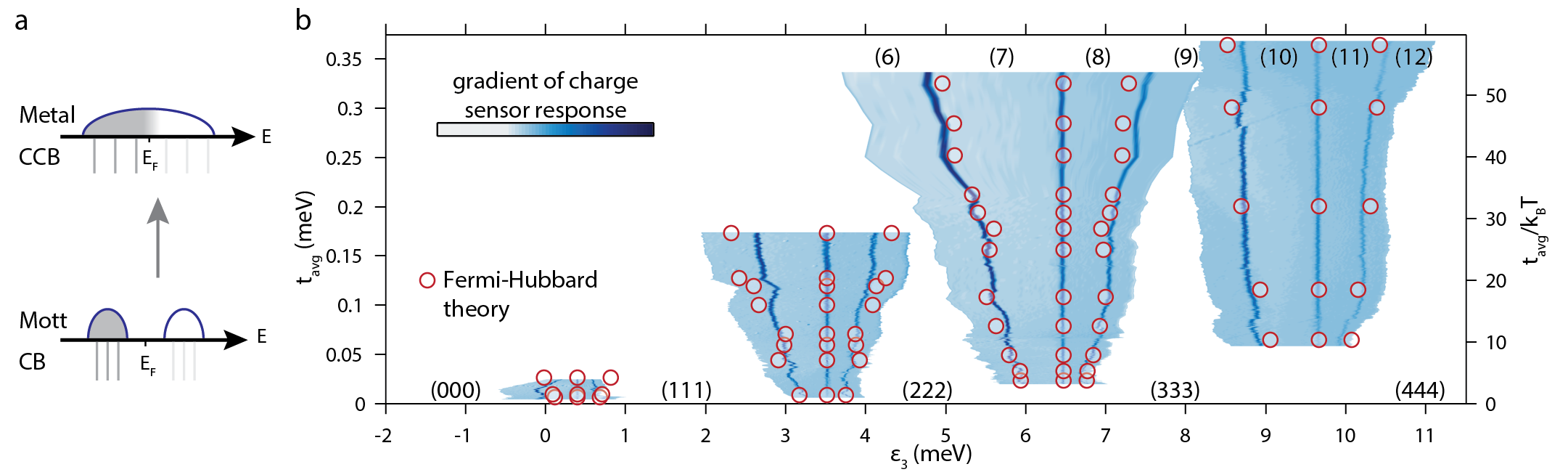}
 	\caption{\textbf{Collective Coulomb blockade physics in the Fermi-Hubbard phase space.} \textbf{a} Schematic representation of the charge addition spectrum of a Mott insulator at half filling and a triple quantum dot array in Coulomb blockade (bottom) and those of a metallic phase at half filling and a triple quantum dot array in collective Coulomb blockade (top).
\textbf{b} The experimentally accessible parameter space of the Fermi-Hubbard model for a triple quantum dot array as a function of electron filling and nearest-neighbour tunnel coupling. Continuous charge sensing measurements following the charging lines are shown, at calibrated gate values where the dots are filled homogeneously (only $\epsilon_3$ values are shown) and the $t_{ij}$'s are set to be roughly equal. Plotted spacings between the bands are set by the Coulomb interaction terms measured at small tunnel coupling. This is justified for tunnel coupling effects as $t<U$ for the entire diagram, but does omit the effect of non-constant Coulomb interaction terms on the overall offset between the bands. The gate layout, which was designed for spin qubit experiments at fillings around one electron per site and tunnel couplings up to several tens of $\upmu$eV, limits the maximum tunnel coupling we can achieve at fillings below two electrons per dot and the minimum tunnel coupling at higher fillings. Red circles indicate extended Hubbard model calculations of the transitions, using locally quantified Hamiltonian parameters (see Supplementary Section V for details), and text in brackets denotes electron filling.}
 	\label{fig:virtgates}
 \end{figure*}

\noindent The experimental phase diagram is mapped out by the independent control over electron filling and tunnel coupling strength and constructed continuously by linear interpolation of gate values in between multiple calibrated points per miniband (\figureref{3b}) where the on-site energies and tunnel couplings are well calibrated and the interaction energies measured (see Supplementary Section V for the data). The main effect of increased nearest-neighbour tunnel coupling on the addition spectrum is indeed a widening of the minibands at the expense of the collective gap at uniform filling, analogous to the reduction of the Mott gap with increasing tunnel coupling.  The gap between minibands continues to decrease with increasing tunnel coupling, but will be prohibited from closing completely by the charging energy of what has essentially become one large dot, a quantity inversely proportional to its large but finite total capacitance. Both regimes can be clearly distinguished using transport measurements through the quantum dot array as well as via charge sensing (see Supplementary Section VI). Note that the experimental tools for Hamiltonian engineering remain valid over the entire measured diagram, as the agreement between measurements and numerical calculation shows. It is thus possible to calibrate and characterize site-specific quantum dot parameters up to large values of tunnel coupling, in this case reaching $U/t = 7.1(4)$. Also, the large energy scales obtained compared to temperature, $t/k_BT = 54(5)$, will mean pure-state initialization comes for free even for significantly scaled-up devices.\\

\section*{\centering Conclusion}

\noindent These results demonstrate an increased level of control over semiconductor quantum dot arrays that allows for the simulation of Fermi-Hubbard physics in the low-temperature, strong-interaction regime (i.e., $k_BT \ll t < U$) where quantum correlations in charge and spin are the strongest. Further automation \cite{Eendebak2016} will be paramount for the efficient calibration of longer chains, which can for instance, together with programmable disorders in on-site energies, be mapped onto the physics of many-body localization \cite{Basko2006}. Advances in device size, connectivity and homogeneity are underway as well in the pursuit of scalable quantum computing, the results of which can be directly leveraged. Examples include scalable gate layouts for 1D arrays  \cite{Medford2013,Zajac2016} as well as square \cite{Thalineau2012} and triangular \cite{Seo2013} geometries, industrial-grade fabrication processes \cite{Intel2014} and magnetically quiet $^{28}$Si substrates \cite{Veldhorst2014}, that open up further possibilities for quantum simulation experiments with quantum dots.\\

\newpage

\subsection*{Acknowledgements}
The authors acknowledge useful discussions with M. Veldhorst, A.F. Otte, R. Sensarma and the members of the Vandersypen group, sample fabrication by F.R. Braakman, set-up preparation by T.A. Baart and experimental assistance from M. Ammerlaan, J. Haanstra, R. Roeleveld, R. Schouten, and R. Vermeulen. This work is supported by the Netherlands Organization of Scientific Research (NWO) VICI program, the European Commission via the integrated project SIQS, the Japan Society for the Promotion of Science (JSPS) Postdoctoral Fellowship for Research Abroad, LPS-MPO-CMTC and the Swiss National Science Foundation.

\subsection*{Author contributions}
T.H., T.F., C.J.D. and L.J. performed the experiment and analysed the data, C.R. and W.W. grew the heterostructure, X.L. and S.D.S. performed the
theoretical analyses with X.L. carrying out the numerical simulations, T.H., T.F., X.L., L.J., S.D.S. and L.M.K.V. contributed to the interpretation of the data, and T.H. wrote the manuscript (X.L. wrote part of the Supplementary Information), with comments from T.F., X.L., S.D.S. and L.M.K.V. 

\subsection*{Author information}
The authors declare no competing financial interests. Correspondence and requests for materials should be addressed to L.M.K.V. (l.m.k.vandersypen@tudelft.nl).

\clearpage
	\newpage
	
	\renewcommand{\figurename}{Figure~S}
	\renewcommand{\tablename}{Table~S}
	\renewcommand{\theequation}{S\arabic{equation}}
	\renewcommand{\thesection}{\Roman{section}} 
	\renewcommand{\thesubsection}{\Alph{subsection}}
	\setcounter{figure}{0}
	
	\pagenumbering{roman}
	
\begin{centering}
{\Large Supplementary Information for} \\ \vspace{0.2cm}
{\Large \textbf{Quantum simulation of a Fermi-Hubbard model using a semiconductor quantum dot array}}\\
\vspace{0.4cm}

{\normalsize T. Hensgens$^{1}$, T. Fujita$^{1}$, L. Janssen$^{1}$, Xiao Li$^{2}$, C.~J. Van Diepen$^{3}$}\\  
{\normalsize C. Reichl$^{4}$, W. Wegscheider$^{4}$, S. Das Sarma$^{2}$, L.~M.~K. Vandersypen$^{1}$}\\

\vspace{0.4cm}
\normalsize{$^{1}$QuTech and Kavli Institute of Nanoscience, TU Delft, 2600 GA Delft, The Netherlands}\\
\normalsize{$^{2}$Condensed Matter Theory Center and Joint Quantum Institute, University of Maryland, College Park, Maryland 20742, USA}\\
\normalsize{$^{3}$QuTech and Netherlands Organization for Applied Scientific Research (TNO), 2600 AD Delft, The Netherlands}\\
\normalsize{$^{4}$Solid State Physics Laboratory, ETH Z\"{u}rich, 8093 Z\"{u}rich, Switzerland}\\
\end{centering}

\vspace{1cm}
\renewcommand{\baselinestretch}{2.0}\normalsize
\tableofcontents
\addtocontents{toc}{~\hfill\textbf{Page}\par}
\renewcommand{\baselinestretch}{1.0}\normalsize
	
\newpage
\section{Methods and materials}
The triple quantum dot sample was fabricated on a GaAs/Al$_{0.25}$Ga$_{0.75}$As heterostructure that was grown by molecular-beam epitaxy. The 85-nm-deep 2D electron gas has an electron density of 2.0 $\times$ 10$^{11}$ cm$^{-2}$ and 4 K mobility of 5.6 $\times$ 10$^6$ cm$^2$V$^{-1}$s$^{-1}$. All sample structures were defined using electron-beam lithography, with metallic gates (Ti/Au) and ohmic contacts (Ni/AuGe/Ni) deposited on the bare wafer in a lift-off process using electron-beam evaporation, similarly to the definition of metallic markers, leads and bonding pads, and with sample mesas defined using a diluted Piranha wet etch. The plunger gates were connected to bias-tees on the printed circuit board, allowing for fast sweeps and RF excitations to be applied in addition to DC voltages. RF reflectometry \cite{Barthel2010a} of the sensing dot channel conductance is done at 110.35 MHz employing a homebuilt LC circuit on the printed circuit board. The sample was cooled down in an Oxford Kelvinox 400HA dilution refridgerator to a base temperature of 45mK whilst applying positive bias voltages to all gates. With the sample cold and the dots formed through application of appropriate voltages to the metallic gates, read-out was performed by feeding the RF reflectometry a roughly -99 dBm carrier wave, the reflected signal of which gets amplified at 4 K and subsequently demodulated and measured using custom electronics. For more detailed methods please see Baart \textit{et. al.} \cite{Baart2015a}.	
	
\section{Theory of classically coupled quantum dots}
\label{sec:filling}

In this section we model charge stability measurements at low tunnel couplings using the picture of a system of classically coupled quantum dots that can exchange particles with an adjacent reservoir. In particular, we would like to explain how one homogeneously fills an inhomogeneous system.

\subsection{Charge addition spectrum}

Isolated quantum dots are well described by a classical capacitance model  \cite{VanderWiel2002a}. This description is valid as long as tunnel coupling energies are negligible compared to capacitive (Coulomb) effects, and if so, the charge states $s$ of the system are simply described by the set of individual dot occupations $(n_1,n_2,..)$ as the $n_i$'s are good quantum numbers. As has been shown previously  \cite{Yang2011a}, one can map the classical capacitance model onto an extended Hubbard model with single-particle energies $\epsilon_i$ for electrons occupying the $i$th dot and effective on-site and inter-site Coulomb interaction terms $U_i$ on the $i$th dot and $V_{ij}$ between the $i$th and $j$th dot, respectively. Using site-and-spin-specific electronic creation and annihilation operators $c^{\dag}_{i\sigma}$ and $c_{i\sigma}$ and dot occupations $n_{i}=\sum_{\sigma}{c^{\dag}_{i\sigma}c_{i\sigma}}$, we get
\begin{align}
H = -\sum_{i}\varepsilon_{i}n_{i}+\sum_{i}\dfrac{U_i}{2}n_i(n_i-1)+\sum_{i,j\neq i}V_{ij}n_{i}n_{j},
\label{Eq:classicalHubbard}
\end{align}
which is readily diagonalized with eigenenergies $E(n_1,n_2,...)=-\sum_{i}\varepsilon_{i}n_{i}+\sum_{i}\dfrac{U_i}{2}n_i(n_i-1)+\sum_{i,j\neq i}V_{ij}n_{i}n_{j}$. Because we experimentally probe changes in the equilibrium charge state of the array as we couple it to  adjacent electron reservoirs, typically kept at an equal and constant electrochemical potential $\mu$ and temperature $k_BT$, we are interested in the charge addition spectrum $\frac{\partial\langle N\rangle}{\partial \mu}$, with 
\begin{align}
	\langle N\rangle = k_BT \dfrac{\partial \ln \mathcal{Z}}{\partial \mu}, \quad \mathcal{Z} = \text{Tr}\{\exp[-(H-\mu N)/k_BT]\}, \label{Eq:PartitionFunction}
\end{align}
where $N=\sum_{i}n_i$ is the total electron number and $\mathcal{Z}$ is the grand partition function. In this classical case and at constant chemical potential $\mu=0$, the equations for the charge addition spectrum $\frac{\partial\langle N\rangle}{\partial \mu}=\frac{\langle N^2\rangle - \langle N\rangle^2}{k_BT} $ simplify to simple Boltzmann-weighted sums over the charge states $s$, namely $\mathcal{Z} = \sum_{s}\exp[-E_s/k_BT]$ and $\langle N^k\rangle = \frac{1}{\mathcal{Z}}\sum_{s}N_s^k\exp[-E_s/k_BT]$. Note that for the purpose of finding the charge transitions, we can ignore any spin-degeneracy of the charges states. The charge stability measurements shown in the main text effectively show two-dimensional slices of the charge addition spectrum as a function of changes in the $\epsilon_i$'s, whose electrostatic control using the physical gate voltages is described in Sup. \ref{sec:virtualgates}.\\

\noindent Note that it is common in the field of quantum dots to draw schematic `ladder' pictures that keep track of charge transitions through the definition of electrochemical potentials $\mu_{ss'} \equiv E(s)-E(s')$  between charge states $s$ and $s'$ that differ by $\Delta n_i=1$. These have the nice features of being linearly sensitive to $\epsilon_i$ and of clearly indicating where the transitions are, as Eq. \eqref{Eq:PartitionFunction} tells us to expect such a transition for $\mu_{ss'}=\mu$, i.e., when the specific electrochemical potential lines up with the reservoir. As such, this is a very popular representation for small systems of well-isolated dots  \cite{VanderWiel2002a}. As a general scheme, though, this picture has several problems. The bookkeeping becomes tedious with a large number of relevant states, for instance in a system with a large number of dots. Furthermore, states $s$ and $s'$ can hybridize in the case of nonzero $t/U$, clouding the representation.

\subsection{Homogeneous filling using linear combinations of site-specific energy offsets}

We can experimentally control the filling of the quantum dot array by changing the energy difference between the electronic states at the Fermi level of the reservoir and those of the dot array itself. The former can be done by applying a bias voltage to the relevant Fermi reservoir, the latter by applying voltages to top gates that influence the single-particle energies $\epsilon_i$ on the dots. Because the partition function is only sensitive to changes in $H-\mu N$, we can equivalently think about changes in the $\epsilon_i$'s as influencing the chemical potential directly through $\delta(\mu N)=\delta(\sum_{i}\epsilon_{i}n_{i})$, which at uniform filling, simplifies to $\delta\mu = \langle \delta\epsilon_i\rangle$. This allows for a different look at the gate control over a quantum dot array with $M$ sites. Instead of thinking about $M$ different $\epsilon_i$'s, we can define one global chemical potential term $\mu'=\langle\epsilon_i\rangle$ and $M-1$ energy differences $\delta_i = \epsilon_i-\mu'$, the latter of which describe the setting of some (controllable) disorder potential landscape at a fixed chemical potential $\mu'$.\\
 
\noindent It leaves the question, however, of which linear set of ${\delta\epsilon_i}$ we choose to change the global $\mu'$. In the case of a large and homogeneous system, the obvious choice would be to change all $\epsilon_i$ equally, as this will uniformly and homogeneously fill all dots in the system. For the triple-quantum-dot sample described in the main text, however, both the finite size (e.g. only one of the three dots has two direct neighbours) and inhomogeneous interaction terms (e.g. $U_1 \neq U_2$) mean a different approach is needed, which we illustrate below.

 \begin{figure*}[h!]
	\centering
 	\includegraphics{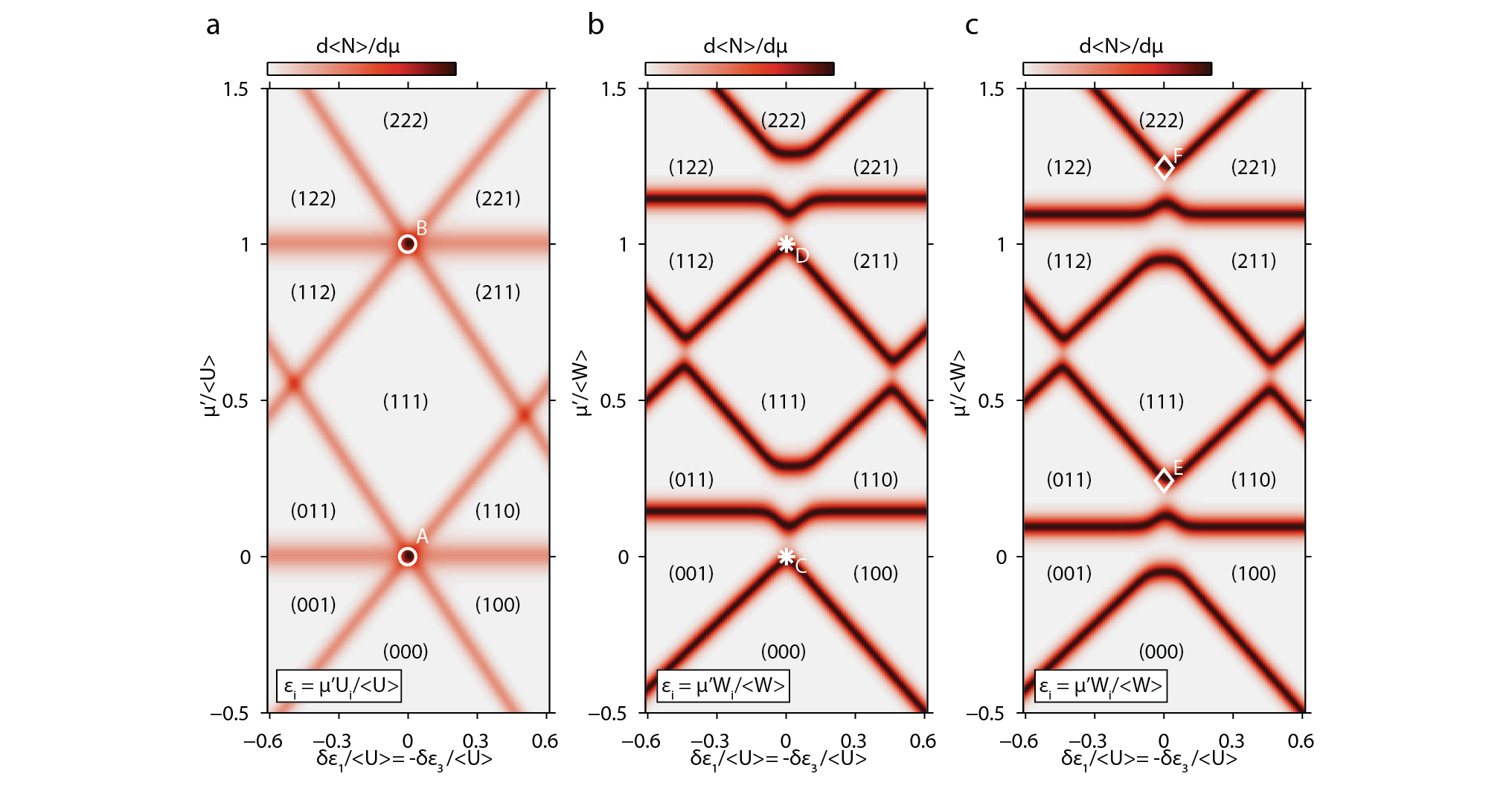}
 	\caption{Simulated charge addition spectra following Eq.~\eqref{Eq:PartitionFunction} for a triple-dot with $U_2 = 1.05 U_1 = 0.95 U_3$ and up to two particles per dot connected to a reservoir at $\mu=0$ and $k_BT=0.02U$ ($>$10 times larger than for the experiments described in the main text), with $V_{ij}=0$ and $\delta_i=0$ \textbf{a} or with $V_{12}=V_{12}=2V_{13}=0.2U$ and $\delta_i=0$ \textbf{b} or $\delta_1=\delta_3=0$ and $\delta_2=U/15$ \textbf{c}. States are denoted by charge occupation ($n_1n_2n_3$) and specific degeneracy points A-F are referred to in the text. The relation between $\epsilon_i$ and $\mu'$ specified in the bottom left boxes applies to the vertical line at zero (horizontal) detuning.}
 	\label{fig:fillingdifs}
 \end{figure*} 

In order to homogeneously fill the three dots, we have to align some well-defined points in the $(\epsilon_1,\epsilon_2,\epsilon_3)$-space. In the case of $V_{ij}=0$, and focussing on the regime from 0 to 2 electrons per site, the only obvious choice would be to identify and align points A --where the eight charge states (000) to (111) are degenerate-- and point B --where (111) to (222) are degenerate-- (see Fig.~S\ref{fig:fillingdifs}a). It shall come as no surprise that these points are lined up by changing the on-site single particle energies by ratio of their on-site repulsions $\epsilon_i=\mu' U_i / \langle U\rangle$. Analogously, under finite $V_{ij}$, we use the ratio of the sum of all locally relevant interaction energies $W_i = U_i+\sum_{j\neq i}V_{ij}$ as $\epsilon_i=\mu' W_i / \langle W\rangle$. Note, however, that the inter-site repulsion breaks particle-hole symmetry and moves states with more than one particle added to a homogeneously filled state to higher energy, meaning we can only find points with at most 4 degenerate states. This leaves us with two obvious choices. Either we align points C --where (000), (100), (010) and (001) are degenerate-- and D --where (111), (211), (121) and (112) are degenerate-- (see Fig.~S\ref{fig:fillingdifs}b), or we align points E --where (110), (101), (110) and (111) are degenerate and F --where (221), (212), (221) and (222) are degenerate-- (see Fig.~S\ref{fig:fillingdifs}c), the two of which are particle-hole partners of the same total state.\\

\begin{figure*}[h!]
	\centering
 	\includegraphics{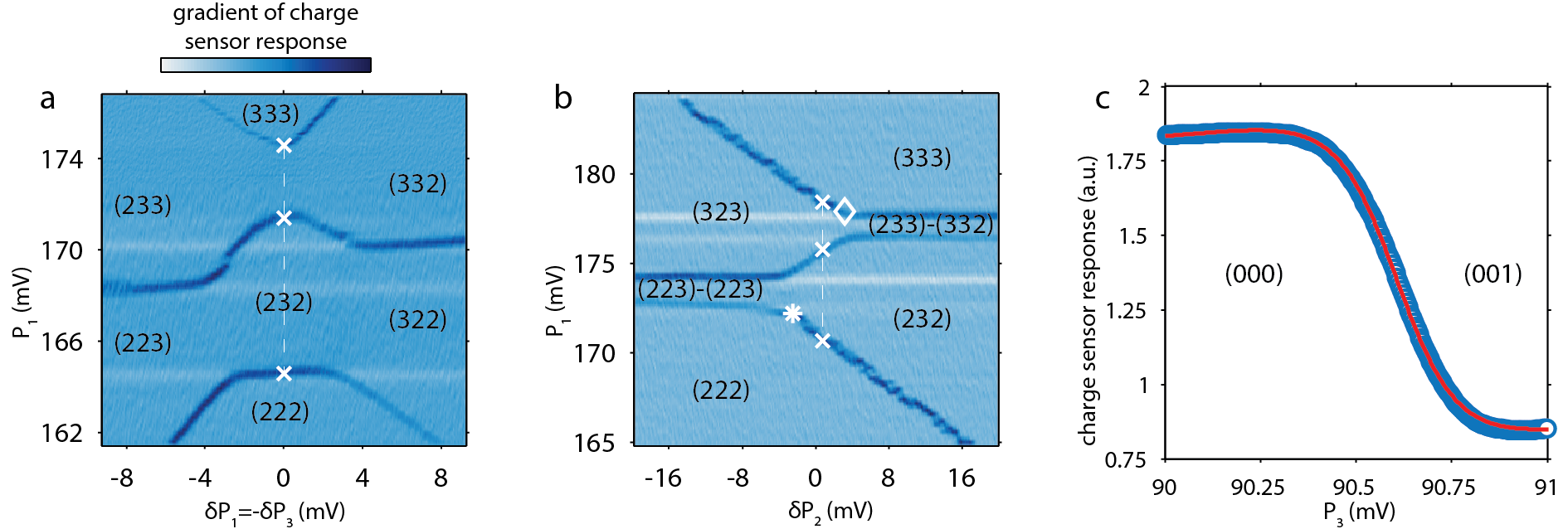}
 	\caption{\textbf{a} Measured charge stability diagrams of the 222-333 miniband as a function of homogeneous filling (only $P_1$ values are shown) and offset in the outer two dot energies by changing $P_1$ and $P_3$ in opposite directions, akin to the simulations of Fig.~S\ref{fig:fillingdifs}c. \textbf{b} Comparable measurement as a function of the offset in the middle dot energy by changing $P_2$. The $P_1$ values are somewhat different from \textbf{a} because these measurements were taken at slightly different tunnel coupling tunings. The white diamond and asterisk indicate (roughly) the position of the same degeneracy points as shown in Fig.~S\ref{fig:fillingdifs}. Note the nonzero background in charge sensor response we find in experiments, which is due to a direct coupling between the swept gate voltages and the sensing dot conductivity. \textbf{c} Broadening of a charge addition line due to the finite temperature of the (rightmost) Fermi reservoir. Fermi-Dirac fit of the transition is shown in red, which, together with the lever arm (see Sup. \ref{sec:PATvsPol}) yields an effective reservoir temperature of 72.2(5) mK.}
 	\label{fig:detuningdat}
 \end{figure*}

\noindent If we define a miniband as the region in chemical potential where we go from one uniform filling to the next (the first miniband is thus the transition region between (000) and (111)), it becomes clear that the inter-site Coulomb terms already widen the miniband at zero tunnel coupling. On top of this, too large a deviation in the site-specific energy offsets $\epsilon_i$'s from the desired values (which amounts to disorder in the dot energies) can also increase the miniband width. Note, however, that the width remains minimized as long as the $\epsilon_i$'s are tuned between the values denoted by the crosses and diamonds of Fig.~S\ref{fig:fillingdifs}. Fig.~S\ref{fig:detuningdat} shows how we can experimentally tune the $\epsilon_i$ to be in this regime, as well as a measurement of the electron reservoir temperature. 

\section{Eliminating cross-capacitances through the definition of virtual gates}
\label{sec:virtualgates}
Gate voltages determine the potential landscape in the 2DEG, which in turn defines the dots and their physical parameters. Each gate is designed with the tuning of a specific parameter in mind: plunger gates $P_i$ are used to change dot single-particle energy offsets $\epsilon_i$ and as such occupation $n_i$, the barrier gates $B_{ij}$ to change tunnel coupling $t_{ij}$ between dots or $\Gamma_{ij}$ to Fermi reservoirs. Of these parameters, the tunnel couplings are the hardest to measure and depend on gate voltages in a sometimes intricate, nonlinear manner. Changes in $\epsilon_i$, however, can be tracked relatively easily by following transitions in the charge stability diagram  and depend only linearly on gate values for voltage changes on the order of several tens of millivolts. In general, small changes in the energy offsets of the three dots will thus be a linear combination of voltage changes on each of the total of seven gates:
\begin{equation}
\left(\begin{array}{c} \delta\epsilon_1 \\ \delta\epsilon_2 \\ \delta\epsilon_3 \\ \end{array}  \right) = 
\left( \begin{array}{ccccccc}
\alpha_{11} & \alpha_{12} & \alpha_{13} & \alpha_{14} & \alpha_{15} & \alpha_{16} & \alpha_{17}  \\
\alpha_{21} & \alpha_{22} & \alpha_{23} & \alpha_{24} & \alpha_{25} & \alpha_{26} & \alpha_{27}  \\
\alpha_{31} & \alpha_{31} & \alpha_{33} & \alpha_{34} & \alpha_{35} & \alpha_{36} & \alpha_{37}  
\end{array} \right) 
\left(\begin{array}{c} \delta P_1 \\ \delta P_2 \\ \delta P_3 \\ \delta B_{1L} \\ \delta B_{12} \\ \delta B_{23} \\ \delta B_{3R} \\ \end{array}\right)
\label{eq:virtgates}
\end{equation}

\noindent Of these 21 matrix elements, the three $\alpha_{ii}$'s describe the coupling of the plungers $P_i$ to the energy offset $\epsilon_i$ of their respective dot $i$. The other 18 elements are simple cross-capacitances, whose values can easily be related to the $\alpha_{ii}$'s through the slope of charge addition lines (Fig.~S\ref{fig:virtgates}a). This leaves the relative weights of the $\alpha_{ii}$'s and the absolute value of one of the elements to be determined. As the difference between the single-particle energies of two dots stays fixed along a polarization line, we can determine the relative weights from the slope of these lines (Fig.~S\ref{fig:virtgates}b). The absolute value of $\alpha_{22}$ can be found using photon-assisted tunneling measurements (see Sup. \ref{sec:PATvsPol}). For the measurements presented in this paper, the matrix has been measured multiple times for different fillings and tunnel couplings: the `plunger' side $\alpha_{11}$-$\alpha_{33}$ of the matrix was measured 25 times in total and the `barrier' part $\alpha_{14}$-$\alpha_{37}$ 12 times (Fig.~S\ref{fig:virtgates}c). In between these points, we used linear interpolation as function of measured tunnel coupling to extract matrix elements when needed.\\

\noindent With all matrix elements known, we can deterministically change $\epsilon_i$'s, a technique which is extensively used throughout the results presented here in two main ways:
\begin{itemize}
\item measuring Hamiltonian parameters through direct interpretation of features in the addition spectrum
\item definition of `virtual gates', both for plunger and barrier gates, that greatly simplify the tuning process
\end{itemize}
 
 \begin{figure*}[h!]
	\centering
 	\includegraphics{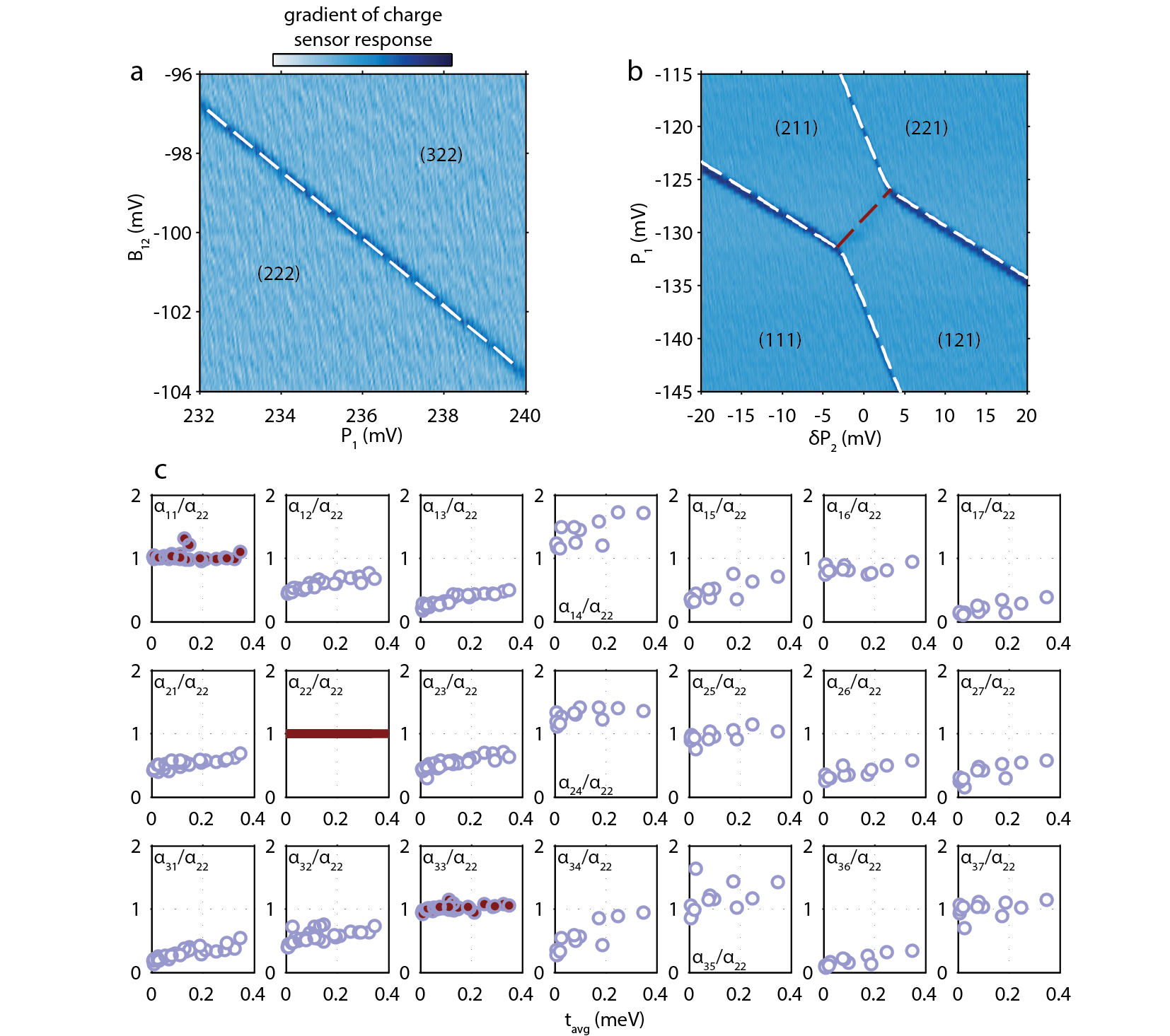}
 	\caption{\textbf{a} Cross-capacitance measurement of gates $P_1$ and $B_{12}$ on the left dot. Dashed white line is fit of the transition. The slope of the charge transition (fit in white) yields the relative strength ($\delta B_{12}/\delta P_1=-\alpha_{11}/\alpha_{14}$) of the two gates in gating the single-particle energy offset of the leftmost dot. \textbf{b} Charge stability diagram showing the anticrossing and polarization line (fit in red) between the left and middle dot, yielding the relative strength $\alpha_{11}=\alpha_{21}+(\delta P_2/\delta P_1)(\alpha_{22}-\alpha_{12})$ of the two plungers in gating their respective dots. Automated edge finding and fitting procedures are outlined in Sup. \ref{sec:fitting}.  \textbf{c} Measured matrix elements $\alpha_{ij}/\alpha_{22}$ as a function of tunnel coupling. No visual distinction is made between the measured matrix elements at different electron filling.}
 	\label{fig:virtgates}
 \end{figure*} 
 
As an example of the last point, consider trying to increase tunnel coupling $t_{12}$. Defining a virtual gate step $\delta B_{12} \rightarrow \delta VB_{12}=(\delta P_1,\delta P_2,\delta P_3,\delta B_{12})$ with $\left( \begin{smallmatrix} \delta P_1 \\ \delta P_2 \\ \delta P_3 \end{smallmatrix} \right) = -\delta B_{12} \left( \begin{smallmatrix} \alpha_{11} & \alpha_{12} & \alpha_{13} \\ \alpha_{21} & \alpha_{22} & \alpha_{23} \\ \alpha_{31} & \alpha_{32} & \alpha_{33} \end{smallmatrix} \right)^{-1} \left( \begin{smallmatrix} \alpha_{14} \\ \alpha_{24} \\ \alpha_{34} \end{smallmatrix} \right)$ allows for making the barrier separating dots 1 and 2 more (or less) transparent without changing the energy offsets $\epsilon_i$ of any of the dots, which is to say, stay at the same location in the charge stability diagram. Analogously, if we want to change the energy offset of one dot only, say the leftmost dot, we can do this using a simple combination of all three plunger gates: $\left( \begin{smallmatrix} \delta P_1 \\ \delta P_2 \\ \delta P_3 \end{smallmatrix} \right) = \left( \begin{smallmatrix} \alpha_{11} & \alpha_{12} & \alpha_{13} \\ \alpha_{21} & \alpha_{22} & \alpha_{23} \\ \alpha_{31} & \alpha_{32} & \alpha_{33} \end{smallmatrix} \right)^{-1} \left( \begin{smallmatrix} \delta \epsilon_1 \\ 0 \\ 0 \end{smallmatrix} \right)$.

\section{Extracting Coulomb terms from the charge-stability diagram}
\label{sec:fitting}

Much of the day-to-day work in quantum dot arrays in general and for the measurements described here in particular consists of the interpretation of features in the charge stability diagram. In the case of well isolated dots with localized electrons ($t/U\ll1$) this essentially boils down to one-dot features (parallel lines) and two-dot features (anticrossings and associated polarization lines). Indeed, pattern recognition of anticrossings is the crucial step in the automated initial tuning of double quantum dots  \cite{Eendebak2016a}.\\

\noindent In general, the processing of a charge stability diagram starts with finding charge transitions in the raw sensor dot data using an edge finding algorithm. The results are filtered to only leave edge sections with more than a threshold number of points. Next, we employ a k-means algorithm to cluster the edges into line sections. Depending on the data, manual input might be needed, either in the selection of relevant clusters or, sometimes, in the case of noisy data, manual selection of points. In determining on-site interaction terms $U_i$, calculating the orthogonal distance between two parallel lines suffices. Equation~\eqref{eq:virtgates} tells us what site-specific energy offset $\epsilon_i$ (and thus energy) difference corresponds to that distance in gate values. In the case of an anti crossing, we employ a 2D fitting routine in a rotated frame $\left( \begin{smallmatrix} y \\ x \end{smallmatrix} \right) = \frac{1}{2} \left( \begin{smallmatrix} \delta\epsilon_i+\delta\epsilon_j \\ \delta\epsilon_i-\delta\epsilon_j \end{smallmatrix} \right) = \left( \begin{smallmatrix} ~~1/2 & ~1/2 \\ -1/2 & ~1/2 \end{smallmatrix} \right) \left( \begin{smallmatrix} \alpha_{ii} & \alpha_{ij} \\ \alpha_{ji} & \alpha_{jj} \end{smallmatrix} \right) \left( \begin{smallmatrix} \delta P_i \\ \delta P_j \end{smallmatrix} \right)$, simultaneously fitting both branches in a least squares sense to $y-y_0=\pm \left( V_{ij}/2+\sqrt{(x-x_0)^2+t_{ij}^2} \right)$. Fitting parameters are three of the matrix elements (corresponding to the angles of the two dot lines and the polarization line), the two offsets $x_0$ and $y_0$ and the two energies $V_{ij}$ and $t_{ij}$. Both the procedures to find $U_i$ and $V_{ij}$ are limited to $t/U<0.15$, as around this value for the tunnel coupling there are no straight line sections in charge addition left where two well-defined localized charge states meet and the specific dots are well defined.

 \begin{figure*}[h!]
 	\centering
 	\includegraphics{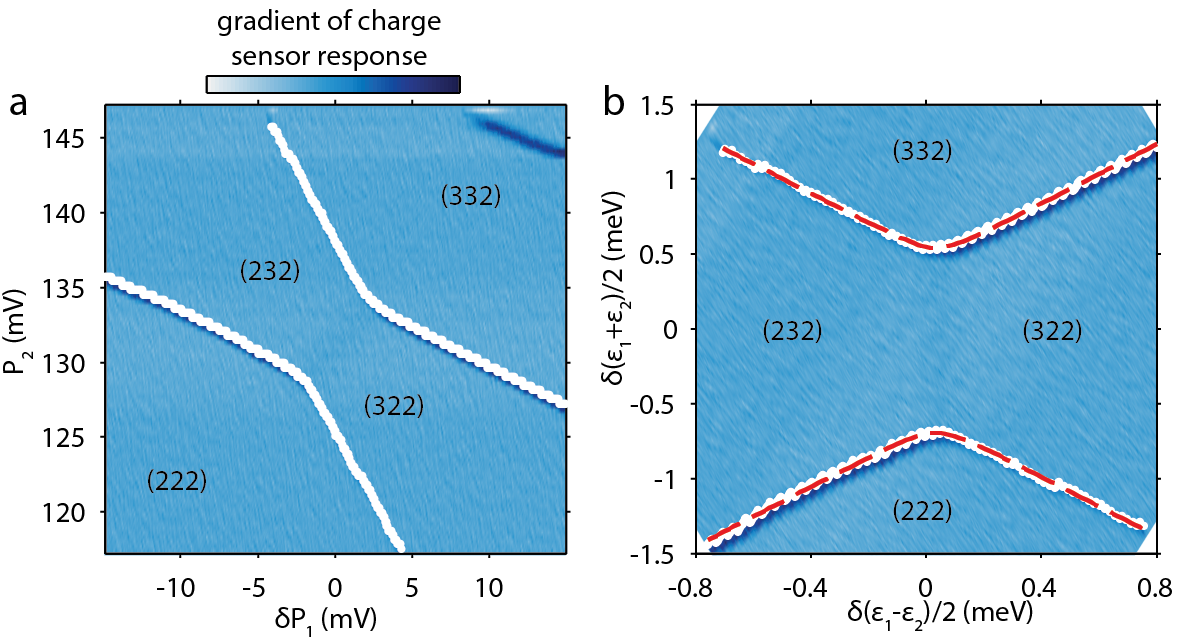}
 	\caption{\textbf{a} Example of an anticrossing measurement between dots 1 and 2 as function of their respective plunger voltages. In white are the points that are detected as lying on the relevant charge transition lines. In the top right corner we see the transition line of dot 3. \textbf{b} The same anticrossing in the rotated frame. Least squares fit is shown in red.}
 	\label{fig:fitaX}
 \end{figure*}

\section{Determination of lever arm and tunnel coupling}
\label{sec:PATvsPol}

Here we explain the methods used for extracting lever arm and tunnel coupling.\\

\begin{figure*}[h!]
 	\centering
 	\includegraphics{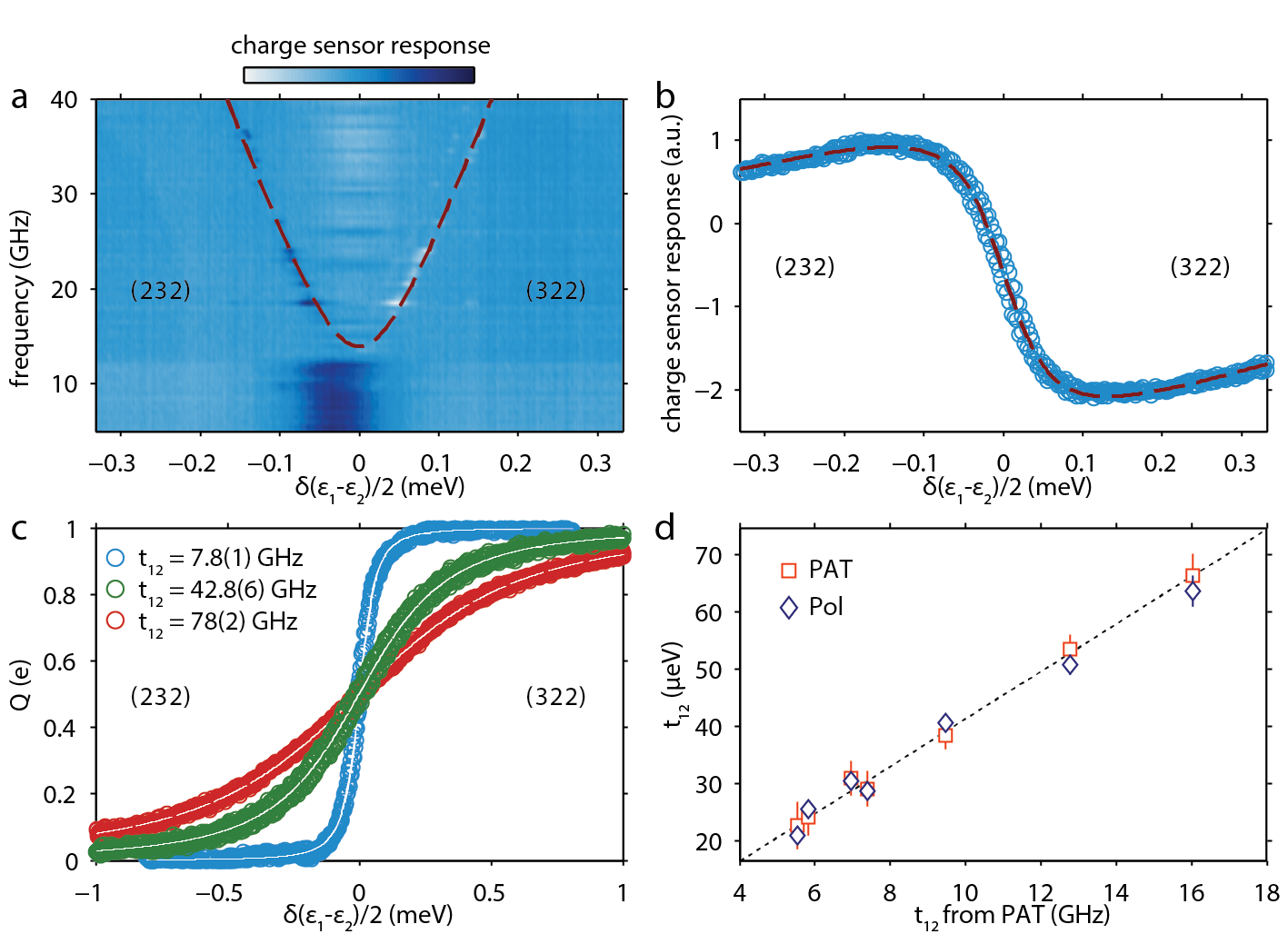}
 	\caption{\textbf{a} Example of a PAT measurement. Plotted is the difference in charge sensor response between applying the microwave excitation or not as a function of detuning. Dashed red line is the fit. \textbf{b} Example of a polarization line width measurement, with fit in red. \textbf{c} Excess charge as function of detuning for three different tunnel couplings, showing that this characterization method works up to significantly larger tunnel couplings than PAT. \textbf{d} Comparison of PAT and polarization line width (Pol) measurements. The data is well explained by assuming a constant lever arm $\alpha_{22}=83(1)~\upmu$eV/mV  between gate $P_2$ and the middle dot. Text in brackets denote relevant charge states, error bars are fit uncertainties.}
 	\label{fig:polvspat}
 \end{figure*}

\noindent At low tunnel frequencies, the measurement method of choice for both  lever arm and tunnel coupling is photon assisted tunneling (PAT, for an example see Fig.~S\ref{fig:polvspat}a). In PAT, as we turn on an AC excitation to gate $P_2$, we check for changes in the charge sensor response, indicating photon assisted tunneling between adjacent dots for certain values of detuning and frequency corresponding to the hybridized charge state spectrum of the double dot  \cite{Oosterkamp_1998aa}. The energy difference between bonding and antibonding states yields the minimum in frequency ($2t$) and the slope away from the transition is described by the lever arm between detuning voltages applied to the gates and single-particle energy difference change between the two dots. The need to generate AC excitations and transmit them without significant losses through coaxial cables in the fridge, however, limits the maximum tunnel frequency we can accurately determine with this method to roughly 20 GHz (83 $\upmu$eV).\\
 
\noindent As an alternative to PAT, one can determine the tunnel coupling by assessing the width of the polarization line  \cite{DiCarlo2004a}. The excess charge (say on the left dot) transition is broadened both by an effective electron temperature and by the tunnel coupling (Fig.~S\ref{fig:polvspat}b). Charge sensor response is however not a direct measurement of excess charge. Not only does there exist a finite cross-capacitance between the gate voltages and the charge sensor response that leads to a finite slope away from the transition, we also found a back-effect of the excess charge on the sensing dot, leading to a different slope on either side of the transition. We fit the data with the following equation, taking this back-effect into account to first order in excess charge:

\begin{equation}
V(\epsilon) = V_0 + \delta V Q(\epsilon) + \left[ \frac{\delta V}{\delta \epsilon}\vert_{Q=0} + \left( \frac{\delta V}{\delta \epsilon}\vert_{Q=1} - \frac{\delta V}{\delta \epsilon}\vert_{Q=0}  \right) Q(\epsilon) \right] \epsilon
\label{eq:polline}
\end{equation}

\noindent where $V(\epsilon)$ is the charge sensor response as a function of the detuning $\epsilon=\delta (\epsilon_i-\epsilon_j)$ away from to the transition and $V_0$, $\delta V$ and $\frac{\delta V}{\delta\epsilon}$ are the background signal, sensitivity and gate-sensor coupling, respectively. Note that $\epsilon$ is a linear combination of the swept gate voltages described by Eq.~\eqref{eq:virtgates}, taking the relevant cross-capacitances and lever arm into account. Excess charge on the left dot is described by $Q(\epsilon)$:

\begin{equation}
Q(\epsilon) = \frac{1}{2} \left( 1 + \frac{\epsilon}{\Omega} \tanh \left( \frac{\Omega}{2\mathrm{k}_\mathrm{B} T_e} \right) \right)
\label{eq:excesscharge}
\end{equation}

\noindent with $\Omega=\sqrt{\epsilon^2+4t_{ij}^2}$ and effective temperature k$_\mathrm{B}T_e \approx 6.5~\upmu$eV (1.6 GHz). Fig S\ref{fig:polvspat}d shows that both methods are in good agreement, and that we can assume a constant lever arm between the middle dot and gate $P_2$. Together with the relative cross-capacitances, this finishes the characterization of the gates and allows for the deterministic sweeping of dot-specific single-particle energies in meV using appropriate gate voltages applied to the top gates.

\section{Extended Hubbard model simulations}
\label{sec:simu}

In this section we provide some theoretical details about the collective Coulomb blockade (CCB) physics as well as its simulation based on an extended Hubbard model. 

\subsection{Collective Coulomb blockade in a simplified Hubbard model}
\begin{figure}[!]
\centering \includegraphics[width=1\textwidth]{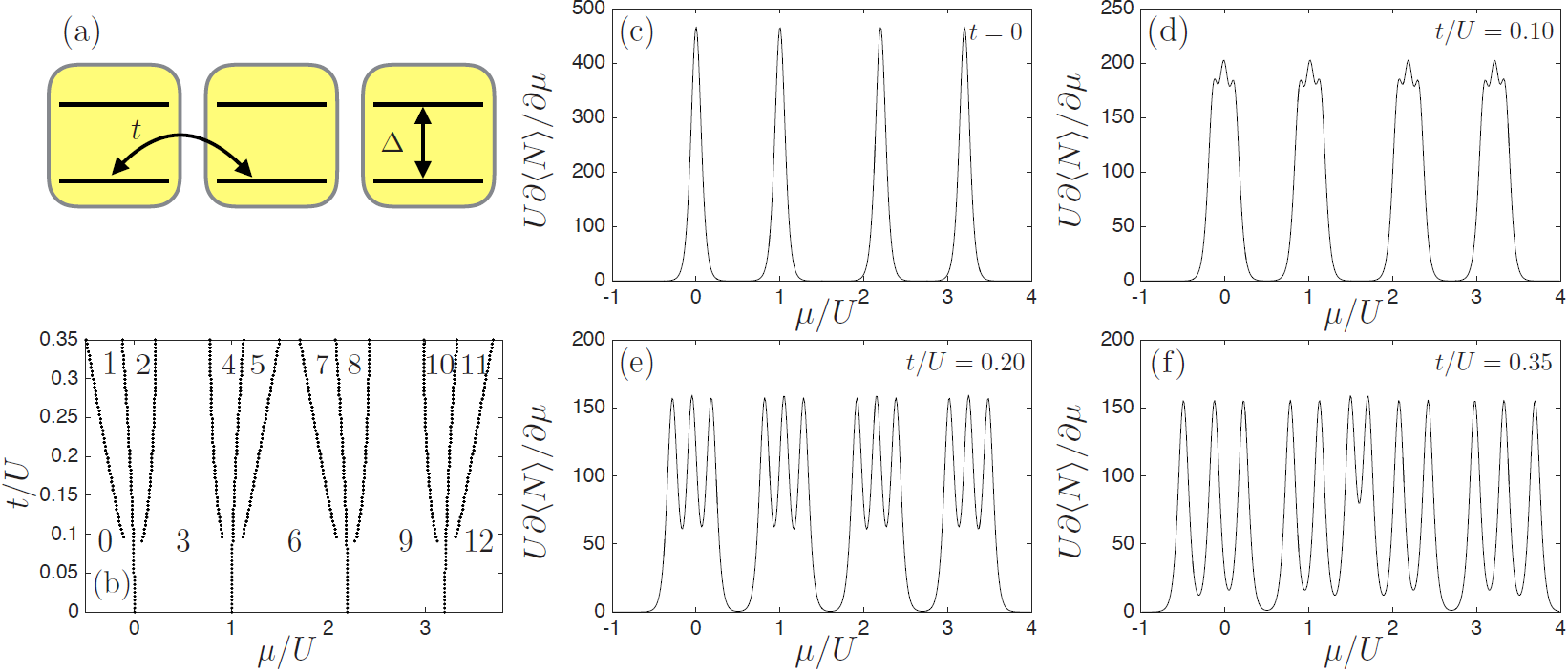}
\caption{\label{Fig:CCB-1} 
\textbf{a} Energy diagram for the simplified Hubbard model [Eq.~\eqref{Eq:SimplifiedHubbard}] in a triple dot system. Each energy level is doubly degenerate due to the spin degrees of freedom. 
\textbf{b} Peaks in the electron addition spectrum for the triple dot system in \textbf{a}. The numbers in the figure indicate the average electron numbers in the system when the chemical potential resides at the respective gap. Here we use $\Delta/U = 0.2$, and $k_BT /U = 0.04$. 
\textbf{c-f} Line cuts for the addition spectrum in \textbf{b} at different values of $t/U$.
}
\end{figure}

It is known that the classical Coulomb blockade effect arises purely from the charging effects of the quantum dots. When electron tunnelling between quantum dots is allowed, however, quantum fluctuations compete with the classical charging effects and give rise to a rich phase diagram, which is known as collective Coulomb blockade~ \cite{Stafford1994a}. 
As we discussed in the main text, there will be three different regimes in this phase diagram: at weak tunnel couplings the quantum dot states split into minibands but the isolated Coulomb blockade of each individual dot is preserved; at intermediate tunnel couplings the Coulomb blockade of individual dots is lost, but the gap between minibands remains open; finally, in the large tunnel coupling limit the gap between minibands can become comparable to temperature, and the system will be in a metallic state. 
To illustrate this physics in more details, we consider the following model, 
\begin{align}
	H = -t\sum_{\langle ij\rangle,\alpha}\left(c^{\dagger}_{i,\alpha}c_{j,\alpha}+\text{H.c.}\right)+\sum_{j,\alpha}\varepsilon_{j,\alpha}c^{\dagger}_{j,\alpha}c_{j,\alpha}+\dfrac{U}{2}\sum_{i}n_i(n_i-1), \label{Eq:SimplifiedHubbard}
\end{align}
which is a simplified version of the actual model we used in the text. Specifically, we have set a uniform tunnel coupling $t$ and Hubbard $U$, while ignoring the inter-site Coulomb interaction term $V_{ij}$. Moreover, $\varepsilon_{j,\alpha}$ is the single-particle energy for an electron in state $\alpha$ of the $j$th dot, and $n_j = \sum_{\alpha}c^{\dagger}_{j,\alpha}c_{j,\alpha}$ is the total number of electrons in the $j$th dot. The schematic energy diagram for a triple dot system is shown in Fig.~S\ref{Fig:CCB-1}a, where each energy level is doubly degenerate due to electron spin.\\ 

\noindent The metal-insulator transition in such a system is best captured by the charge addition spectrum \eqref{Eq:PartitionFunction}, and a typical example is shown in Fig.~\ref{Fig:CCB-1}b, which is precisely what we measure in the experiment (Fig. 3a in the main text). 
At $t=0$ we can see that there are four critical chemical potentials $\mu$ at which electrons can be added to the triple dot (Fig.~S\ref{Fig:CCB-1}c). For the present model, these four peaks occur at $\mu = 0$, $U$, $2U+\Delta$, and $3U+\Delta$, respectively. Each peak is triply degenerate, as the energy cost to add electrons to any of the three dots is identical. 
For nonzero but small tunnel couplings (Fig.~S\ref{Fig:CCB-1}b-c) each triply degenerate peak at $t=0$ starts to split into minibands, indicating the breakdown of Coulomb blockade in each dot. However, different minibands are still separated by gaps that arise from a collective origin, reminiscent of the energy gap in a Mott insulator. 
Finally, at sufficiently high tunnel couplings the gap between different minibands closes (Fig.~S\ref{Fig:CCB-1}f), and Coulomb blockade is overwhelmed by temperature altogether.\\

\noindent However, there are several important differences between this model and what we measured in the experiment. 
First of all, in the present model we ignored the inter-site Coulomb interactions $V_{ij}n_in_j$, which will split the peaks in the addition spectrum even at zero tunnel coupling, as discussed in Sup. \ref{sec:filling}. 
Second, because it is difficult to experimentally fix the absolute chemical potential over large areas of the parameter space due to nonlinearities in the gating effect, the addition spectrum in the main text was constructed by lining up the middle transition within each miniband, and measuring the chemical potentials of adjacent transitions with respect to those. As we can see from Fig.~S\ref{Fig:CCB-1}b, such an approximation is justified at small $t/U(<0.15)$, although it neglects any tunnel coupling dependence on the interaction terms.
Finally, as also discussed in Sup. \ref{sec:filling}, it requires an inhomogeneous change in the site-specific energy offsets to homogeneously fill the array, that cannot be captured by a single chemical potential term. We take this into account in a manner akin to that in Sup. \ref{sec:filling}, namely by replacing $\mu N$ by $\sum_{i}\epsilon_in_i$ in the term $H - \mu N$ in the partition function. The `global' chemical potential $\mu$ can be obtained by equating both terms: $\mu = \frac{1}{N}\sum_{i}\epsilon_in_i$. 

\subsection{Characterizing the model parameters}
We now describe in detail how we extract the model parameters from the experimental measurements. 
In particular, we will use the estimate of $U_{i}$ as an example to demonstrate that the way the parameters $t_{ij}$, $U_i$, and $V_{ij}$ are extracted in the main text is valid in the small tunnel coupling limit ($t/U<0.15$) that this experiment covers.\\ 

\noindent For convenience, we rewrite the model Hamiltonian in the main text here, i.e., 
\begin{align}
	H = - \sum_i{\epsilon_i n_i} - \sum_{\langle i,j\rangle,\sigma}{t_{ij}(c^{\dag}_{i\sigma}c_{j\sigma}+\mathrm{H.c.})} + \sum_{i}{\frac{U_i}{2}n_i(n_i-1)} + \sum_{i,j}{V_{ij}n_in_j}, \label{Eq:Hubbarmodel}
\end{align}
To begin with, we note that the eigenstates of the above Hamiltonian in the $t=0$ limit can be obtained exactly, as the ground states of the triple dot system can be represented by the exact charge state $(n_1, n_2, n_3)$, where $n_i$ denotes the electron occupation in the $i$th dot (see I).  
In this regime, one can show that on the $\epsilon_2$-$\epsilon_3$ plane the border between the $(111)/(112)$ region and the border between the $(111)/(110)$ region are exactly separated by an energy of $U_3$. Similarly, the border between the $(111)/(121)$ region and the border between the $(111)/(101)$ region are separated by an energy of $U_2$. This is exactly the distance indicated by the vertical arrow in Fig. 2c in the main text.\\

\begin{figure}[!]
\centering \includegraphics[width=0.75\textwidth]{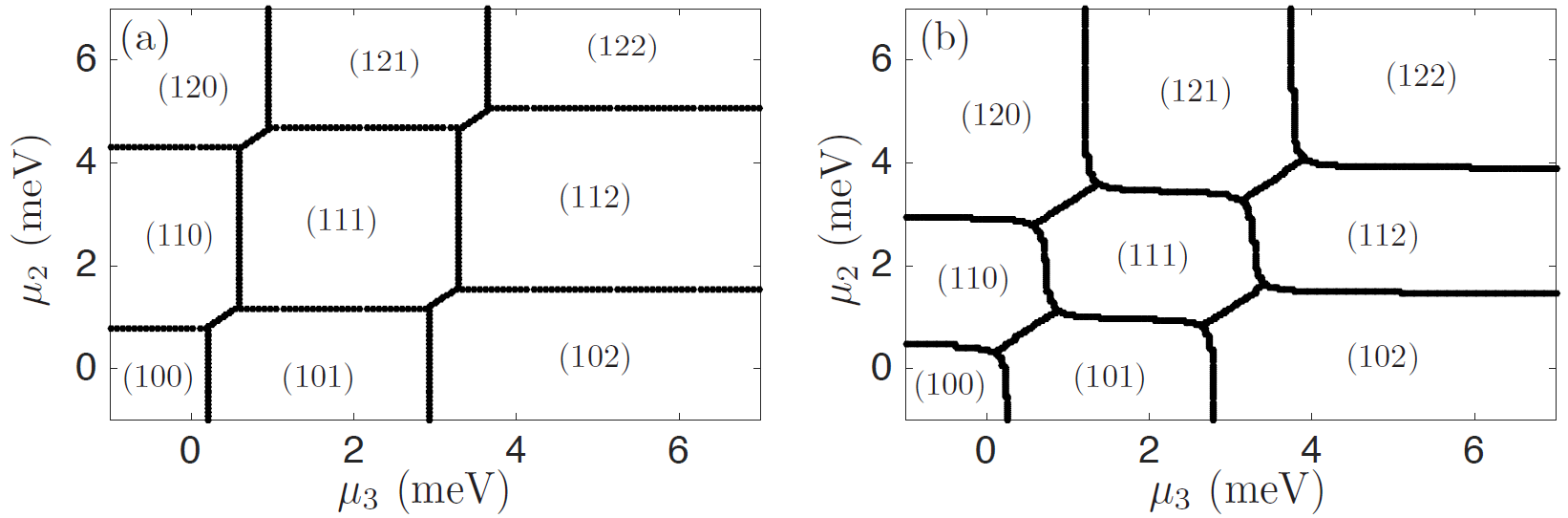}
\caption{\label{Fig:CCB-2} 
\textbf{a} Charge stability diagram for a triple dot system with parameters given in Eq.~\eqref{Eq:TripleDotA}.
\textbf{b} Charge stability diagram for a triple dot system with parameters given in Eq.~\eqref{Eq:TripleDotB}. The ground state in the region labeled by `X' is (100) in both figures. 
}
\end{figure}

\noindent In the presence of a nonzero but small tunnel coupling, we expect that such an estimate is still reasonable. For example, consider a triple dot system characterized by the following parameters (all in meV). 
\begin{align}
	t = 0.0061,\; U_1 = 3.9815,\; U_2 = 3.4899,\; U_3 = 2.7080,\; V_{12} = 0.4119,\; V_{23} = 0.3527,\; V_{13} = 0.1125, \label{Eq:TripleDotA}
\end{align}
Its charge stability diagram is given in Fig.~S\ref{Fig:CCB-2}a, which is qualitatively the same as the one shown in Fig.~2c in the main text. Now that the tunnel coupling is nonzero, the ground state of the system is no longer an exact charge state $(n_1, n_2, n_3)$, but generally a superposition of different charge states. To retain a connection to the $t=0$ limit, we keep labelling sections of the charge stability as $(n_1, n_2, n_3)$, but with the distinction in mind that $(n_1,n_2,n_3)$ no longer denotes the exact ground state, but instead the charge state with the largest weight in the actual ground state.
Because the tunnel coupling is vanishingly small ($t/U\simeq 0.002$) in Fig.~2c, the charge stability diagram in the $t=0$ limit will be all but the same as Fig.~S\ref{Fig:CCB-2}a. 
Moreover, we can determine the values of $U_2$ and $U_3$ using the method described in the previous paragraph and find that $U_2 = \SI{3.4463}{meV}$ and $U_3=\SI{2.7130}{meV}$, respectively, which is reasonably close to the corresponding model parameters in Eq.~\eqref{Eq:TripleDotA}. \\

\noindent In fact, for the range of tunnel couplings available in this experiment, such an approach to extract model parameters remains valid. 
For example, consider a triple dot characterized by the following parameters (in meV): 
\begin{align}
	t = 0.1770,\; U_1 = 2.920,\; U_2 = 2.390,\; U_3 = 2.530,\; V_{12} = 0.550,\; V_{23} = 0.470,\; V_{13} = 0.270, \label{Eq:TripleDotB}
\end{align}
which has a much larger tunnel coupling. The corresponding charge stability diagram is shown in Fig.~S\ref{Fig:CCB-2}b. We find that the structure of the charge stability diagram remains qualitatively the same as that in Fig.~S\ref{Fig:CCB-2}a, and if we again extract the values of $U_2$ and $U_3$ using the same method, we find that $U_2 = \SI{2.48}{meV}$ and $U_{3}=\SI{2.56}{meV}$, which still agrees reasonably well with the original model parameters in Eq.~\eqref{Eq:TripleDotB}.\\ 

\noindent Granted, at sufficiently large $t/U$ the structure of the charge stability diagram will change drastically, and the present method to extract model parameters is bound to fail. However, as we never enter those regimes, our fitting method serves the purpose of this experiment.

\begin{figure}[!]
\includegraphics[scale=1]{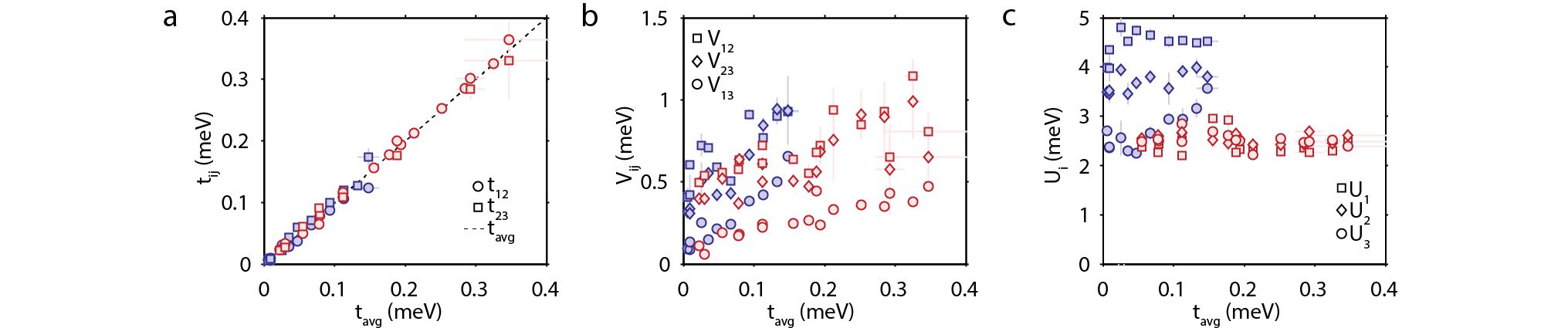}
\caption{\label{Fig:energies} 
\textbf{a} Calibrated tunnel couplings and measured inter-site Coulomb \textbf{b} and on-site Coulomb \textbf{c} terms at calibrated values of the average tunnel coupling, corresponding to the experimental parameter space plot shown in Fig. 3b of the main text. Error bars are fit errors. Blue fill indicates data from the first subband from 0 to 6 electrons, red fill data from the second subband from 6 to 12 electrons.
}
\end{figure}

\subsection{Measuring the width of each miniband}
Having fit the model parameters from the measurements, we now turn to the width of minibands in the electron addition spectrum, which is a key quantity in collective Coulomb blockade physics. 
The comparison between the theory and experiment requires careful consideration in this work. In particular, Fig.~S\ref{Fig:CCB-1}b cannot be measured directly in this experiment, because of the site-specific energy offsets $\epsilon_i$ required to maintain homogeneous filling in the system, and because the absolute value of the chemical potential cannot be calibrated across a wide range.\\ 

\noindent In order to achieve the best comparison, in the simulation we draw each of the twelve charge addition lines according to the experimental procedure as follows: (i) When the system has $N = 3n$ electrons, its ground state is tuned to be the $(n,n,n)$ state; (ii) the charge addition line that separates $N=3n$ particles and $N=3n\pm1$ particles is constructed by identifying critical points at which the four states $(n,n,n)$, $(n\pm1,n,n)$, $(n,n\pm1,n)$, and $(n,n,n\pm1)$ are degenerate. Such a procedure allows us to obtain for each charge addition line a curve in the three-dimensional parameter space spanned by $(\epsilon_1, \epsilon_2, \epsilon_3)$. 
Subsequently we are able to obtain three different widths for each miniband, one along each $\epsilon_i$ direction. Fig. 3b in the main text shows the charge addition lines as a function of $\epsilon_3$. \\

\begin{table}[!]
\centering
\caption{Transition points for a triplet dot system with parameters given in Eq.~\eqref{Eq:TripleDotC}. The label $n_1\to n_2$ indicates that this data is for the transition from a total of $n_1$ particles to $n_2$ particles. $\epsilon_{i}$ ($i=1,2,3$) are the `local' chemical potentials on each dot, while $\mu$ is the `uniform' chemical potential as defined in Eq.~\eqref{Eq:PartitionFunction}. The last two columns compare the experimental and theoretical total width of the fourth miniband. All energies are given in meV. \label{Table:CCB-1}}
\begin{tabular}{c|cccccc|cc}
\hline\hline{}
Transition & $6\to7$ & $7\to8$ & $8\to9$ & $9\to10$ & $10\to11$ & $11\to12$ & width of $9\to12$ (Theory) & (Exp.)\\
\hline
$\epsilon_1$ & 6.3800 & 7.1280 & 8.0460 & 9.7800  & 10.5620 & 11.5140 & 1.7340 & 1.7553\\
$\epsilon_2$ & 7.4600 & 8.3664 & 9.4788 & 11.5800 & 12.5276 & 13.6812 & 2.1012 & 2.0699\\
$\epsilon_3$ & 6.6600 & 7.4432 & 8.4044 & 10.2200 & 11.0388 & 12.0356 & 1.8156 & 1.6494\\
\hline
$\mu$   & 6.6200 & 7.3400 & 8.9800 & 10.1000 & 11.0200 & 13.2600 & 3.1600 & - \\
\hline\hline
\end{tabular}
\end{table}

\begin{table}[!]
\centering
\caption{Width of the fourth miniband in Fig. 3b in the main text. All energies are in meV. \label{Table:CCB-2}}
\begin{tabular}{c|cc|cc|cc}
\hline\hline
$t$ & $\epsilon_1$ (Th.) & $\epsilon_1$ (Exp.) & $\epsilon_2$ (Th.) & $\epsilon_2$ (Exp.) & $\epsilon_3$ (Th.) & $\epsilon_3$ (Exp.)\\
\hline
0.0777 & 1.0032 & 1.0012 & 1.1616  & 1.1279	& 1.0164 & 0.9813 \\ 
0.1120 & 1.1552 & 1.0918 & 1.4288  & 1.4623 & 1.2312 & 1.4011 \\
0.1880 & 1.5824 & 1.5420 & 1.8216  & 1.7554 & 1.6192 & 1.4832 \\ 
0.2930 & 1.7340 & 1.7553 & 2.1012  & 2.0699 & 1.8156 & 1.6494 \\
0.3480 & 2.0352 & 2.1758 & 2.3532  & 2.4338 & 1.9080 & 1.8251 \\ 
\hline\hline
\end{tabular}
\end{table}

\noindent Next we provide some additional details regarding the comparison of miniband width shown in Fig. 3b in the main text. We use the following set of quantum dot parameters as an example (all in meV): 
\begin{align}
	t = 0.2930,\; U_1 = 2.260,\; U_2 = 2.770,\; U_3 = 2.480,\; V_{12} = 0.650,\; V_{23} = 0.570,\; V_{13} = 0.430, \label{Eq:TripleDotC}
\end{align}
which corresponds to the data with the second largest tunnel couplings in the fourth miniband in Fig. 3b in the main text. 
First of all, it is helpful to show the `uniform' chemical potential $\mu$ that correspond to the specific $\epsilon_i$'s (as defined in Eq.~\eqref{Eq:Hubbarmodel}). Such a comparison is shown in Table~S~\ref{Table:CCB-1}. We can see that in the three-dimensional parameter space the charge addition line defined by $(\epsilon_1, \epsilon_2, \epsilon_3)$ can be very different from the one defined by $(\mu, \mu, \mu)$. This shows that the distinction is important, and a simple simulation with a uniform chemical potential as in Fig.~S\ref{Fig:CCB-1}b will not compare well with the experiment. 
Second, note that the simulations are done for the specific middle dot detuning denoted by the asterisk in Fig. S\ref{fig:fillingdifs}b and Fig. S\ref{fig:detuningdat}b, whereas the experimental detuning will be in between that situation and the detuning denoted by the diamond in Fig. S\ref{fig:fillingdifs}c and Fig. S\ref{fig:detuningdat}b. This means that although the total width of the miniband will be fixed, the relative position of the middle transition between the outer transitions (which we denote $\alpha$ and which will be close to 0.5) depends on the specific middle dot detuning. To overlay the simulation results on the experimental data in Fig. 3b of the main text, we used values of $\alpha = (0.5,0.6,0.65,0.6)$ for the four minibands, respectively.
Finally, Table~S~\ref{Table:CCB-2} gives an overview for the width of the fourth miniband at different tunnel couplings, as Fig. 3b in the main text only plots the data along the $\epsilon_3$ direction. It can be seen that the theory compares well with the experiment along all three directions, which further corroborates the consistency of our measurements. \\ 

\section{Isolated versus collective Coulomb blockade in charge and transport}
\label{sec:transport}

\begin{figure}[H]
\includegraphics[scale=1]{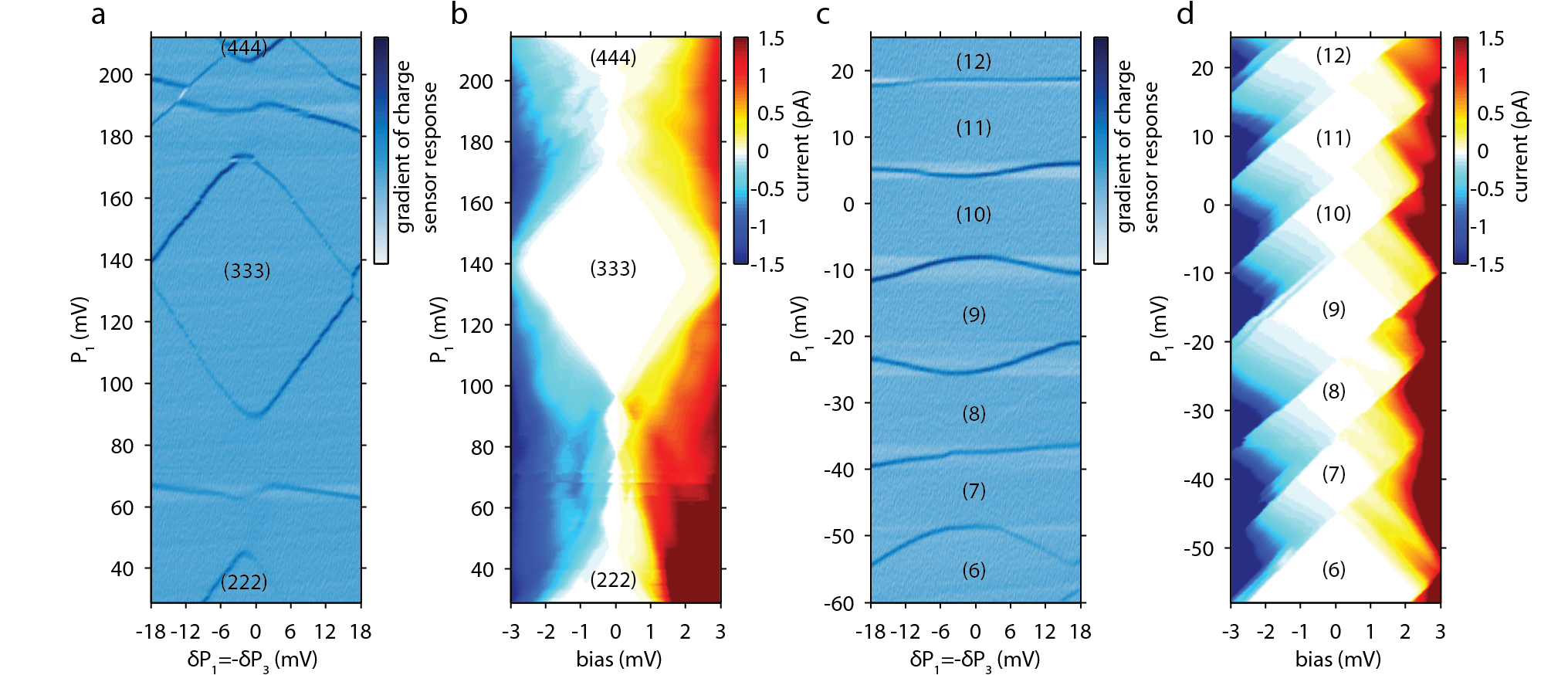}
\caption{\label{Fig:transport} 
\textbf{a} \& \textbf{c} Charge stability diagram around the (333) regime in the low and high tunnel coupling regimes, respectively, using a combination of all seven gates (only $P_1$ values are shown) that sweeps the local fillings equally. \textbf{b} Transport through the array following the zero-detuning line of Fig. 2b of the main text as a function of applied bias (60\% on leftmost and 40\% on bottom right reservoir). In the (333) state, this applied bias has to overcome the local (strong) Coulomb repulsion in order for current to flow. \textbf{d} Similar data in the high tunnel coupling regime. Whereas the individual nature of the dots is all but gone, global (weaker) Coulomb repulsion still prohibits transport at small bias, as expected for the collective Coulomb blockade phase. 
}
\end{figure}

To further investigate the distinct phases, we focus on the regime with around nine electrons in total, corresponding to half-filling of the second band, and look at both charge sensing and transport (Fig. S\ref{Fig:transport}).

\noindent In the localized phase ($t/U < 0.02$), the charge stability diagram shows transition lines following three distinct, well-defined directions, corresponding to the filling of the separate lithographically defined dots. In the delocalized phase ($t/U > 0.15$), this distinct nature is all but lost, highlighting the incipient formation of a large single dot. The same effect can also be seen in transport measurements, as we observe Coulomb diamond sizes as a function of filling. In the localized phase at half filling, local interactions prevent a current from flowing for bias voltages below the local addition energy, similar to a Mott insulator whose Fermi energy resides inside the gap. Adjacent Coulomb diamonds correspond to a Fermi-level inside the miniband and are significantly smaller, allowing current to flow at much smaller bias voltages. In the delocalized phase, however, the notion of a large gap at half-filling is gone, and it is but the charging energy of the entire system that prohibits transport to occur, regardless of filling. The dots are in collective Coulomb blockade, and its transport characteristics are similar to that of a small, metallic island.


\newpage



\begin{thebibliography}{10}
\expandafter\ifx\csname url\endcsname\relax
\def\url#1{\texttt{#1}}\fi
\expandafter\ifx\csname urlprefix\endcsname\relax\def\urlprefix{URL }\fi
\providecommand{\bibinfo}[2]{#2}
\providecommand{\eprint}[2][]{\url{#2}}



\bibitem{Imada1998}
\bibinfo{author}{Imada, M.}, \bibinfo{author}{Fujimori, A.} \& \bibinfo{author}{Tokura, Y.}
\newblock \bibinfo{title}{{Metal-insulator transitions}}.
\newblock \emph{\bibinfo{journal}{Reviews of Modern Physics}}
\textbf{\bibinfo{volume}{70}}, \bibinfo{pages}{1039} (\bibinfo{year}{1998}).

\bibitem{Lee2006}
\bibinfo{author}{Lee, P.~A.}, \bibinfo{author}{Nagaosa, N.} \&
\bibinfo{author}{Wen, X.}
\newblock \bibinfo{title}{{Doping a Mott insulator: Physics of high-temperature superconductivity}}.
\newblock \emph{\bibinfo{journal}{Reviews of Modern Physics}}
\textbf{\bibinfo{volume}{78}}, \bibinfo{pages}{17--85}
(\bibinfo{year}{2006}).

\bibitem{Balents2010}
\bibinfo{author}{Balents, L.}
\newblock \bibinfo{title}{{Spin liquids in frustrated magnets}}.
\newblock \emph{\bibinfo{journal}{Nature}} \textbf{\bibinfo{volume}{464}},
\bibinfo{pages}{199--208} (\bibinfo{year}{2010}).

\bibitem{Anderson2013}
\bibinfo{author}{Anderson, P.~W.}
\newblock \bibinfo{title}{{Twenty-Five Years of High-Temperature Superconductivity - A Personal Review}}.
\newblock \emph{\bibinfo{journal}{JPCS}} \textbf{\bibinfo{volume}{449}},
\bibinfo{pages}{012001} (\bibinfo{year}{2013}).

\bibitem{Joerdens2008}
\bibinfo{author}{J\"{o}rdens, R.}, \bibinfo{author}{Strohmaier, N.}, \bibinfo{author}{G\"{u}nter, K.}, \bibinfo{author}{Moritz, H.} \& \bibinfo{author}{Esslinger, T.}
\newblock \bibinfo{title}{{A Mott insulator of fermionic atoms in an optical lattice}}.
\newblock \emph{\bibinfo{journal}{Nature}} \textbf{\bibinfo{volume}{455}},
\bibinfo{pages}{204--207} (\bibinfo{year}{2008}).

\bibitem{Cirac2012}
\bibinfo{author}{Cirac, J.~I.} \& \bibinfo{author}{Zoller, P.}
\newblock \bibinfo{title}{{Goals and opportunities in quantum simulation}}.
\newblock \emph{\bibinfo{journal}{Nature Phys.}}
\textbf{\bibinfo{volume}{8}}, \bibinfo{pages}{264--266} (\bibinfo{year}{2012}).

\bibitem{Tanese2013}
\bibinfo{author}{Tanese, D.}, \emph{et~al.}
\newblock \bibinfo{title}{{Polariton condensation in solitonic gap states in a one-dimensional periodic potential}}.
\newblock \emph{\bibinfo{journal}{Nat. Commun.}} \textbf{\bibinfo{volume}{4}},
\bibinfo{pages}{1749} (\bibinfo{year}{2013}).

\bibitem{Parsons2016}
\bibinfo{author}{Parsons, M.~F.}, \emph{et~al.}
\newblock \bibinfo{title}{{Site-resolved measurement of the spin-correlation function in the Fermi-Hubbard model}}.
\newblock \emph{\bibinfo{journal}{Science}}
\textbf{\bibinfo{volume}{353}}, \bibinfo{pages}{1253--1256} (\bibinfo{year}{2016}).

\bibitem{Boll2016}
\bibinfo{author}{Boll, M.}, \emph{et~al.}
\newblock \bibinfo{title}{{Spin- and density-resolved microscopy of antiferromagnetic correlations in Fermi-Hubbard chains}}.
\newblock \emph{\bibinfo{journal}{Science}}
\textbf{\bibinfo{volume}{353}}, \bibinfo{pages}{1257--1260} (\bibinfo{year}{2016}).

\bibitem{Cheuk2016}
\bibinfo{author}{Cheuk, L.~W.}, \emph{et~al.}
\newblock \bibinfo{title}{{Observation of spatial charge and spin correlations in the 2D Fermi-Hubbard model}}.
\newblock \emph{\bibinfo{journal}{Science}}
\textbf{\bibinfo{volume}{353}}, \bibinfo{pages}{1260--1264} (\bibinfo{year}{2016}).

\bibitem{Mazurenko2016}
\bibinfo{author}{Mazurenko, A.}, \emph{et~al.}
\newblock \bibinfo{title}{{Experimental realization of a long-range antiferromagnet in the Hubbard model with ultracold atoms}}.
\newblock Preprint at http://arxiv.org/abs/1612.08436 (\bibinfo{year}{2016}).

\bibitem{Manousakis2002}
\bibinfo{author}{Manousakis, E.}
\newblock \bibinfo{title}{{A Quantum-Dot Array as Model for Copper-Oxide Superconductors: A Dedicated Quantum Simulator for the Many-Fermion Problem}}.
\newblock \emph{\bibinfo{journal}{J. Low Temp. Phys.}}
\textbf{\bibinfo{volume}{126}}, \bibinfo{pages}{1501--1513} (\bibinfo{year}{2002}).

\bibitem{Gaudreau2006}
\bibinfo{author}{Gaudreau, L.}, \emph{et~al.}
\newblock \bibinfo{title}{{Stability diagram of a Few-Electron Triple Dot}}.
\newblock \emph{\bibinfo{journal}{Physical Review Letters}}
\textbf{\bibinfo{volume}{97}}, \bibinfo{pages}{036807} (\bibinfo{year}{2006}).

\bibitem{Byrnes2008}
\bibinfo{author}{Byrnes, T.}, \bibinfo{author}{Kim, N.~Y.}, \bibinfo{author}{Kusudo, K.} \& \bibinfo{author}{Yamamoto, Y.}
\newblock \bibinfo{title}{{Quantum simulation of Fermi-Hubbard models in semiconductor quantum-dot arrays}}.
\newblock \emph{\bibinfo{journal}{Physical Review B}}
\textbf{\bibinfo{volume}{78}}, \bibinfo{pages}{075320} (\bibinfo{year}{2008}).

\bibitem{Yang2011}
\bibinfo{author}{Yang, S.}, \bibinfo{author}{Wang, X.} \& \bibinfo{author}{Das Sarma, S.}
\newblock \bibinfo{title}{{Generic Hubbard model description of semiconductor quantum-dot spin qubits}}.
\newblock \emph{\bibinfo{journal}{Physical Review B}}
\textbf{\bibinfo{volume}{83}}, \bibinfo{pages}{161301} (\bibinfo{year}{2011}).

\bibitem{Barthelemy2013}
\bibinfo{author}{Barthelemy, P.} \& \bibinfo{author}{Vandersypen, L.~M.~K.}
\newblock \bibinfo{title}{{Quantum Dot Systems: A versatile platform for
		quantum simulations}}.
\newblock \emph{\bibinfo{journal}{Annalen der Physik}}
\textbf{\bibinfo{volume}{525}}, \bibinfo{pages}{808--826}
(\bibinfo{year}{2013}).

\bibitem{Loss1998a}
\bibinfo{author}{Loss, D.} \& \bibinfo{author}{DiVincenzo, D.~P.}
\newblock \bibinfo{title}{{Quantum computation with quantum dots}}.
\newblock \emph{\bibinfo{journal}{Physical Review A}}
\textbf{\bibinfo{volume}{57}}, \bibinfo{pages}{120} (\bibinfo{year}{1998}).

\bibitem{Petta2005}
\bibinfo{author}{Petta, J.~R.} \emph{et~al.}
\newblock \bibinfo{title}{{Coherent Manipulation of Coupled Electron Spins in Semiconductor Quantum Dots}}.
\newblock \emph{\bibinfo{journal}{Science}}
\textbf{\bibinfo{volume}{309}}, \bibinfo{pages}{2180--2184}
(\bibinfo{year}{2005}).

\bibitem{Koppens2006}
\bibinfo{author}{Koppens, F.~H.~L.} \emph{et~al.}
\newblock \bibinfo{title}{{Driven coherent oscillations of a single electron spin in a quantum dot}}.
\newblock \emph{\bibinfo{journal}{Nature}}
\textbf{\bibinfo{volume}{442}}, \bibinfo{pages}{766--771}
(\bibinfo{year}{2006}).

\bibitem{Petersson2010}
\bibinfo{author}{Petersson, K.~D.}, \bibinfo{author}{Petta, J.~R.}, \bibinfo{author}{Lu, H.} \& \bibinfo{author}{Gossard, A.~C.}
\newblock \bibinfo{title}{{Quantum Coherence in a One-Electron Semiconductor Charge Qubit}}.
\newblock \emph{\bibinfo{journal}{Physical Review Letters}}
\textbf{\bibinfo{volume}{105}}, \bibinfo{pages}{246804} (\bibinfo{year}{2010}).

\bibitem{Baart2015}
\bibinfo{author}{Baart, T.~A.} \emph{et~al.}
\newblock \bibinfo{title}{{Single-spin CCD}}.
\newblock \emph{\bibinfo{journal}{Nature Nanotechnology}}
\textbf{\bibinfo{volume}{11}}, \bibinfo{pages}{330--334}
(\bibinfo{year}{2016}).

\bibitem{Martins2016}
\bibinfo{author}{Martins, F.}, \emph{et~al.}
\newblock \bibinfo{title}{{Noise Suppression Using Symmetric Exchange Gates in Spin Qubits}}.
\newblock \emph{\bibinfo{journal}{Physical Review Letters}}
\textbf{\bibinfo{volume}{116}}, \bibinfo{pages}{116801} (\bibinfo{year}{2016}).

\bibitem{Reed2016}
\bibinfo{author}{Reed, M.~D.} \emph{et~al.}
\newblock \bibinfo{title}{{Reduced Sensitivity to Charge Noise in Semiconductor Spin Qubits via Symmetric Operation}}.
\newblock \emph{\bibinfo{journal}{Physical Review Letters}}
\textbf{\bibinfo{volume}{116}}, \bibinfo{pages}{110402} (\bibinfo{year}{2016}).

\bibitem{Singha2011}
\bibinfo{author}{Singha, A.}, \emph{et~al.}
\newblock \bibinfo{title}{{Two-dimensional Mott-Hubbard electrons in an artificial honeycomb lattice}}.
\newblock \emph{\bibinfo{journal}{Science}}
\textbf{\bibinfo{volume}{332}}, \bibinfo{pages}{1176--1179} (\bibinfo{year}{2011}).

\bibitem{Salfi2016}
\bibinfo{author}{Salfi, J.}, \emph{et~al.}
\newblock \bibinfo{title}{{Quantum simulation of the Hubbard model with dopant atoms in silicon}}.
\newblock \emph{\bibinfo{journal}{Nat. Commun.}}
\textbf{\bibinfo{volume}{7}}, \bibinfo{pages}{11342} (\bibinfo{year}{2016}).

\bibitem{Stafford1994}
\bibinfo{author}{Stafford, C.~A.} \& \bibinfo{author}{Das Sarma, S.}
\newblock \bibinfo{title}{{Collective Coulomb blockade in an array of quantum dots: A Mott-Hubbard approach}}.
\newblock \emph{\bibinfo{journal}{Physical Review Letters}}
\textbf{\bibinfo{volume}{72}}, \bibinfo{pages}{3590} (\bibinfo{year}{1994}).

\bibitem{Farooq2015}
\bibinfo{author}{Farooq, U.}, \bibinfo{author}{Bayat, A.}, \bibinfo{author}{Mancini, S.} \& \bibinfo{author}{Bose, S.}
\newblock \bibinfo{title}{{Adiabatic many-body state preparation and information transfer in quantum dot arrays}}.
\newblock \emph{\bibinfo{journal}{Physical Review B}}
\textbf{\bibinfo{volume}{91}}, \bibinfo{pages}{134303}
(\bibinfo{year}{2015}).

\bibitem{Stehlik2015}
\bibinfo{author}{Stehlik, J.}, \emph{et~al.} 
\newblock \bibinfo{title}{{Fast Charge Sensing of a Cavity-Coupled Double Quantum Dot Using a Josephson Parametric Amplifier}}.
\newblock \emph{\bibinfo{journal}{Phys. Rev. Applied}}
\textbf{\bibinfo{volume}{4}}, \bibinfo{pages}{014018}
(\bibinfo{year}{2015}).

\bibitem{Kouwenhoven1990}
\bibinfo{author}{Kouwenhoven, L.}, \emph{et~al.}
\newblock \bibinfo{title}{{Transport through a finite one-dimensional crystal}}.
\newblock \emph{\bibinfo{journal}{Physical Review Letters}}
\textbf{\bibinfo{volume}{65}}, \bibinfo{pages}{361} (\bibinfo{year}{1990}).

\bibitem{Livermore1996}
\bibinfo{author}{Livermore, C.}, \emph{et~al.}
\newblock \bibinfo{title}{{The Coulomb Blockade in Coupled Quantum Dots}}.
\newblock \emph{\bibinfo{journal}{Science}}
\textbf{\bibinfo{volume}{274}}, \bibinfo{pages}{1332} (\bibinfo{year}{1996}).

\bibitem{Lee2000}
\bibinfo{author}{Lee, S.~D.}, \emph{et~al.}
\newblock \bibinfo{title}{{Single-electron spectroscopy in a triple-dot system: Role of interdot electron-electron interactions}}.
\newblock \emph{\bibinfo{journal}{Physical Review B}}
\textbf{\bibinfo{volume}{62}}, \bibinfo{pages}{7735} (\bibinfo{year}{2000}).

\bibitem{Wei2016}
\bibinfo{author}{Wei, W.}, \emph{et~al.}
\newblock \bibinfo{title}{{Observation of Collective Coulomb Blockade in a Gate-controlled Linear Quantum-dot Array}}.
\newblock Preprint at http://arxiv.org/abs/1603.04625 (\bibinfo{year}{2016}).

\bibitem{Wang2011}
\bibinfo{author}{Wang, X.}, \bibinfo{author}{Yang, S.} \& \bibinfo{author}{Das Sarma, S.}
\newblock \bibinfo{title}{{Quantum theory of the charge-stability diagram of semiconductor double-quantum-dot systems}}.
\newblock \emph{\bibinfo{journal}{Physical Review B}}
\textbf{\bibinfo{volume}{84}}, \bibinfo{pages}{115301}
(\bibinfo{year}{2011}).

\bibitem{Oosterkamp1998}
\bibinfo{author}{Oosterkamp, T.~H.}, \emph{et~al.}
\newblock \bibinfo{title}{{Microwave spectroscopy of a quantum-dot molecule}}.
\newblock \emph{\bibinfo{journal}{Nature}} \textbf{\bibinfo{volume}{395}},
\bibinfo{pages}{873--876} (\bibinfo{year}{1998}).

\bibitem{Eendebak2016}
\bibinfo{author}{Baart, T.~A.}, \bibinfo{author}{Eendebak, P.~T.}, \bibinfo{author}{Reichl, C.}, \bibinfo{author}{Wegscheider, W.} \& \bibinfo{author}{Vandersypen, L.~M.~K.}
\newblock \bibinfo{title}{{Computer-automated tuning of semiconductor double quantum dots into the single-electron regime}}.
\newblock \emph{\bibinfo{journal}{APL}} \textbf{\bibinfo{volume}{108}},
\bibinfo{pages}{213104} (\bibinfo{year}{2016}).

\bibitem{Basko2006}
\bibinfo{author}{Basko, D.~M.}, \bibinfo{author}{Aleiner, I.~L.} \& \bibinfo{author}{Altshuler, B.~L.}
\newblock \bibinfo{title}{{Metal-insulator transition in a weakly interacting many-electron system with localized single-particle states}}.
\newblock \emph{\bibinfo{journal}{Ann. Phys.}} \textbf{\bibinfo{volume}{321}},
\bibinfo{pages}{1126--1205} (\bibinfo{year}{2006}).

\bibitem{Medford2013}
\bibinfo{author}{Medford, J.}, \emph{et~al.}
\newblock \bibinfo{title}{{Self-consistent measurement and state tomography of an exchange-only spin qubit}}.
\newblock \emph{\bibinfo{journal}{Nature Nanotechnology}}
\textbf{\bibinfo{volume}{8}}, \bibinfo{pages}{654--659}
(\bibinfo{year}{2013}).

\bibitem{Zajac2016}
\bibinfo{author}{Zajac, D.~M.}, \bibinfo{author}{Hazard, T.~M.}, \bibinfo{author}{Mi, X.}, \bibinfo{author}{Nielsen, E.}, \& \bibinfo{author}{Petta, J.~R.}
\newblock \bibinfo{title}{{Scalable Gate Architecture for a One-Dimensional Array of Semiconductor Spin Qubits}}.
\newblock \emph{\bibinfo{journal}{Phys. Rev. Applied}}
\textbf{\bibinfo{volume}{6}}, \bibinfo{pages}{054013}
(\bibinfo{year}{2016}).

\bibitem{Thalineau2012}
\bibinfo{author}{Thalineau, R.}, \emph{et~al.}
\newblock \bibinfo{title}{{A few-electron quadruple quantum dot in a closed loop}}.
\newblock \emph{\bibinfo{journal}{Appl. Phys. Lett.}}
\textbf{\bibinfo{volume}{101}}, \bibinfo{pages}{103102} (\bibinfo{year}{2012}).

\bibitem{Seo2013}
\bibinfo{author}{Seo, M.}, \emph{et~al.}
\newblock \bibinfo{title}{{Charge Frustration in a Triangular Triple Quantum Dot}}.
\newblock \emph{\bibinfo{journal}{Physical Review Letters}}
\textbf{\bibinfo{volume}{110}}, \bibinfo{pages}{046803} (\bibinfo{year}{2013}).

\bibitem{Intel2014}
\bibinfo{author}{Natarajan, S.} \emph{et~al.}
\newblock \bibinfo{title}{{A 14nm logic technology featuring 2$^{\mathrm{nd}}$-generation FinFET, air-gapped interconnects, self-aligned double patterning and a 0.0588 $\upmu$m$^2$ SRAM cell size}}.
\newblock \emph{\bibinfo{journal}{2014 IEEE International Electron Devices Meeting}}
\textbf{\bibinfo{volume}{83}}, \bibinfo{pages}{3.7.1--3.7.3} (\bibinfo{year}{2014}).

\bibitem{Veldhorst2014}
\bibinfo{author}{Veldhorst, M.}, \emph{et~al.}
\newblock \bibinfo{title}{{An addressable quantum dot qubit with fault-tolerant control-fidelity}}.
\newblock \emph{\bibinfo{journal}{Nature Nanotechnol.}}
\textbf{\bibinfo{volume}{9}}, \bibinfo{pages}{981--985} (\bibinfo{year}{2014})



\end{thebibliography}

\newpage
\begin{thebibliography}{1}
	\expandafter\ifx\csname url\endcsname\relax
	\def\url#1{\texttt{#1}}\fi
	\expandafter\ifx\csname urlprefix\endcsname\relax\def\urlprefix{URL }\fi
	\providecommand{\bibinfo}[2]{#2}
	\providecommand{\eprint}[2][]{\url{#2}}
	
\bibitem{Barthel2010a}
\bibinfo{author}{Barthel, C.} \emph{et~al.}
\newblock \bibinfo{title}{{Fast sensing of double-dot charge arrangement and
		spin state with a radio-frequency sensor quantum dot}}.
\newblock \emph{\bibinfo{journal}{Physical Review B}}
\textbf{\bibinfo{volume}{81}}, \bibinfo{pages}{161308}
(\bibinfo{year}{2010}).	

\bibitem{Baart2015a}
\bibinfo{author}{Baart, T.~A.} \emph{et~al.}
\newblock \bibinfo{title}{{Single-spin CCD}}.
\newblock \emph{\bibinfo{journal}{Nature Nanotechnology}}
\textbf{\bibinfo{volume}{11}}, \bibinfo{pages}{330--334}
(\bibinfo{year}{2016}).
	
	\bibitem{VanderWiel2002a}
	\bibinfo{author}{Van der Wiel, W.~G.} \emph{et~al.}
\newblock \bibinfo{title}{{Electron transport through double quantum dots}}.
\newblock \emph{\bibinfo{journal}{Rev. Mod. Phys.}}
\textbf{\bibinfo{volume}{75}}, \bibinfo{pages}{1--22} (\bibinfo{year}{2002}).

	
	\bibitem{Yang2011a}
\bibinfo{author}{Yang, S.}, \bibinfo{author}{Wang, X.} \& \bibinfo{author}{Das Sarma, S.}
\newblock \bibinfo{title}{{Generic Hubbard model description of semiconductor quantum-dot spin qubits}}.
\newblock \emph{\bibinfo{journal}{Physical Review B}}
\textbf{\bibinfo{volume}{83}}, \bibinfo{pages}{161301} (\bibinfo{year}{2011}).
		
	\bibitem{Oosterkamp_1998aa}
	\bibinfo{author}{Oosterkamp, T.~H.} \emph{et~al.}
	\newblock \bibinfo{title}{{Microwave spectroscopy of a quantum-dot molecule}}.
	\newblock \emph{\bibinfo{journal}{Nature}} \textbf{\bibinfo{volume}{395}},
	\bibinfo{pages}{873--876} (\bibinfo{year}{1998}).
	
	\bibitem{DiCarlo2004a}
	\bibinfo{author}{DiCarlo, L.} \emph{et~al.}
	\newblock \bibinfo{title}{{Differential Charge Sensing and Charge Delocalization in a Tunable Double Quantum Dot}}.
	\newblock \emph{\bibinfo{journal}{Phys. Rev. Lett.}} \textbf{\bibinfo{volume}{92}},
	\bibinfo{pages}{226801} (\bibinfo{year}{2004}).

	\bibitem{Eendebak2016a}
	\bibinfo{author}{Baart, T. A.} \emph{et~al.}
	\newblock \bibinfo{title}{{Computer-automated tuning of semiconductor double quantum dots into the single-electron regime}}.
	\newblock \emph{\bibinfo{journal}{APL}} \textbf{\bibinfo{volume}{108}},
	\bibinfo{pages}{213104} (\bibinfo{year}{2016}).

	\bibitem{Stafford1994a}
	C. A. Stafford and S. Das Sarma, {Collective Coulomb blockade in an array of quantum dots: A Mott-Hubbard approach}. Phys. Rev. Lett. \textbf{72}, 3590 (1994).
		
\end{thebibliography}
\end{document}